\documentclass[runningheads]{llncs}
\usepackage{graphicx}
\usepackage{xcolor}
\usepackage{hyperref}
\hypersetup{
    colorlinks=true,
    linkcolor=blue,
    filecolor=magenta,      
    urlcolor=cyan,
}
\usepackage{enumitem}

\usepackage{url}

\newcommand{\eg}{\textit{e}.\textit{g}.,~}
\newcommand{\ie}{\textit{i}.\textit{e}.,~}

\begin{document}
\title{Assisted music creation with Flow Machines: towards new categories of new}
%
%
\author{Fran\c{c}ois Pachet, Pierre Roy, Benoit Carr\'e}
\authorrunning{F. Pachet}
%
\institute{CTRL, Spotify
\email{francois@spotify.com} }
\maketitle              

\begin{abstract}This chapter reflects on about 10 years of research in AI-assisted music composition, in particular during the Flow Machines project. We reflect on the motivations for such a project, its background, its main results and impact, both technological and musical, several years after its completion. We conclude with a proposal for new categories of ``new'', created by the many uses of AI techniques to generate novel material.
\end{abstract}

\section{Background and Motivations}

The dream of using machines to compose music automatically has long been a subject of investigation, by musicians and scientists. 

Since the 60s, many researchers used virtually all existing artificial intelligence techniques at hand to solve music generation problems. However, little convincing music was produced with these technologies.

A  landmark result in machine music generation is the \emph{Illiac Suite}, released to the public in 1956~\cite{lejaren:illiac:book:1959}. This piece showed that Markov chains of a rudimentary species (first order, augmented with basic generate-and-test methods) could be used to produce interesting music. We invite the reader to listen to the \href{https://www.youtube.com/watch?v=n0njBFLQSk8}{piece}, composed more than 70 years ago, to appreciate its enduring innovative character.

However, the technology developed for that occasion lacked many fundamental features, to make it actually useable for concrete, professional musical projects. Notably, the experiment involved generate-and-test methods to satisfy various constraints imposed by the authors. Also, the low order of the Markov chain did not produce convincing style imitation. In spite of these many weaknesses, the Illiac suite remains today a remarkable music piece, that can still be listened to with interest.

The Flow Machines project\footnote{conducted at Sony CSL and Sorbonne Universit\'e (2012-2017), and funded by an ERC advanced grant}, aimed at addressing the core technical issues at stake when generating sequences in a given style. In some sense, it addressed the two main weaknesses of the Markov chains used in the \emph{Illiac Suite}: the low order (and the poor style imitation quality) and the controllability, \ie the capacity to force generated sequences to satisfy various criteria, not captured by Markov models.

\subsection{The Continuator}

More precisely, the main motivation for Flow Machines stemmed from the Continuator project. The Continuator~\cite{pachet:continuator:jnmr:2003} was the first interactive system to enable real-time music dialogues with a Markov model. The project was quite successful in the research community, and led to two main threads of investigation: jazz and music education. Jazz experiments were conducted notably with Gy{\"o}rgy Kurtag Jr and Bernard Lubat, leading to various concerts at the Uzeste festival (2000-2004) and many insights concerning the issues related to control~\cite{pachet:03k}.


The education experiments consisted in studying how these free-form interactions with a machine learning component could be exploited for early age music education. Promising initial experiments~\cite{DBLP:journals/cie/PachetA04,addessi_pachet_2005} led to an ambitious project (the \href{http://mirorproject.eu/}{Miror project}) about so-called ``reflexive interactions''. During this project, the Continuator system was  improved and extended, to handle various types of simple constraints. An interesting variant of the Continuator for music composition, called MIROR-Impro was designed, deployed and tested, with which children could generate fully-fledged music compositions, built from  music  generated from their own doodling~\cite{MirorBook}. It was shown also that children could clearly recognize their own style in the material generated by the system~\cite{Khatchatourov2016ActionII}, a  property considered as fundamental for achieving reflexive interaction.

\section{Markov Constraints: Main Scientific Results}
These promising results in the investigation of Markov sequence control led to the Flow Machines project. Technically, most of the work consisted in exploring many types of interesting constraints to be enforced on finite length sequences generated from various machine learning models, such as variable order Markov models. Other tools were developed to offer musicians a comprehensive tool palette with which they could freely explore various creative use cases of style imitation techniques.

\subsection{The ``Markov + X'' roadmap}
\label{section:MarkovPlusXRoadmap}

Markov models are used everywhere, from economics to Google ranking algorithms, and are good at capturing local properties of temporal sequences and abstract them into well-known mathematical concepts (\eg, transition matrice or graphs). A Markov model can easily be estimated from a corpus of sequences in a given style to represent information about how musical notes follow each other. This model can then be used to produce new sequences that will sound more or less similar to the initial sequences. Generation, or sampling, is extremely simple and consists in so-called random walks (also called drunken walks): starting from a random state, transitions are drawn randomly from the model to build a sequence step by step.
However, the remarkable simplicity of random walks in Markov models meets its limitations as soon as one tries to impose specific constraints on the generated sequences. The difficulty arises when one wants to impose simple properties that cannot be expressed as local transition probabilities. For instance, imposing that the last note equals the first one, or that the notes follow some pattern, or that the total duration be fixed in advance. Even more difficult, how to impose that the generated melody is nicely “balanced”, for instance exhibiting a power law distribution of frequencies, characteristic of natural phenomena?

The initial idea was to use the powerful techniques of combinatorial optimization, constraint satisfaction in particular (CP), precisely to represent Markovianity, so that other, additional properties could also be stated as constraints. Indeed, the main advantage of constraint programming is that constraints can be added at will to the problem definition, without changing the solver, at least in principle.

The idea of representing Markovianity as a global constraint was first detailed in~\cite{pachet2011constraints}.
We reformulated Markov generation as a constrained sequence generation problem, an idea which enabled us to produce remarkable examples.
For instance, the \emph{AllDifferent} constraint~\cite{Regin:1994:FAC:199288.178024}, added to a Markov constraint could produce our \emph{Boulez Blues}: a Blues chord sequence in the style of Charlie Parker so that all chords are different! Table~\ref{table:boulezblues} shows the chord progression of the Boulez Blues (check this \href{https://www.francoispachet.fr/wp-content/uploads/2020/04/boulez_blues.mp3}{rendering with jazz musicians}). 

\begin{table}
\begin{center}
\begin{tabular}{ |c|c|c|c| } 
\hline
C7 / Fmin & Bb7 / Ebmin & Ab7 / Db7 & Dbmin / Cmin \\ 
\hline
F7 / Bbmin & Eb7 / Abmin & Gmin / Gbmin & B7 / Gb7 \\ 
\hline
Bmin / E7 & Amin / D7 & Emin / A7 & Dmin / G7  \\ 
\hline
\end{tabular}
\caption{\label{table:boulezblues}The chord sequence of the \href{https://www.francoispachet.fr/wp-content/uploads/2020/04/boulez_blues.mp3}{Boulez Blues}.}
\end{center}
\end{table}

However, this approach was costly, and we did not propose any boundaries on the worst-case complexity. So we started to look for efficient solutions for specific cases.

\subsection{Positional constraints}
\label{section:positionalConstraints}

The first result we obtained was to generate efficiently Markov sequences with positional constraints (called \emph{unary constraints} at the time). Enforcing specific values at specific positions in a finite-length sequence generated from a Markov model turned out to be solvable in polynomial time. The result described in~\cite{pachet:markov:constraints:ijcai:2011} consists in first propagating all the zero probability events (filtering) and then back-propagating probabilities to ensure unbiased sampling. This result enabled us to implement an enhanced version of Continuator and triggered the development of the Flow Machines project.

An interesting and unexpected application of this result (possibility to add positional constraints to Continuator) enabled us to introduce several types of continuations, depending on positional constraints posted on the beginning and ending of a musical phrase. Unary constraints could be used to represent various musical \emph{intentions}, when producing a melody from a Markov model and an input melody provided in real-time.
For instance, we defined the following types of melodic outputs:
\begin{enumerate}
\item Continuation: input is continued to produce a sequence of the same size. A constraint is posted on the last note to ensure that it is “terminal”, \ie occurred at the end of an input
melody, to produce a coherent ending.
\item Variation: is generated by adding two unary constraints that the first and last notes should be the same, respectively, as the first and last notes of the input.
\item Answer: is like a Continuation, but the last note should be the same as the first input note. This creates a phrase that resolves to the beginning, producing a sense of closure.
\end{enumerate}

Additionally, for all types of responses, unary constraints were posted on each intermediary note stating that they should not be initial nor final, to avoid false starts/ends within the melody. This use of positional constraints to bias the nature of a continuation nicely echoes the notion of conversation advocated by composer Hans Zimmer in~\cite{ZimmerConversation}. These constraints substantially improved the musical quality of the responses generated by Continuator. 

\begin{figure}[htp!]\centering
    \includegraphics[width=.9\textwidth]{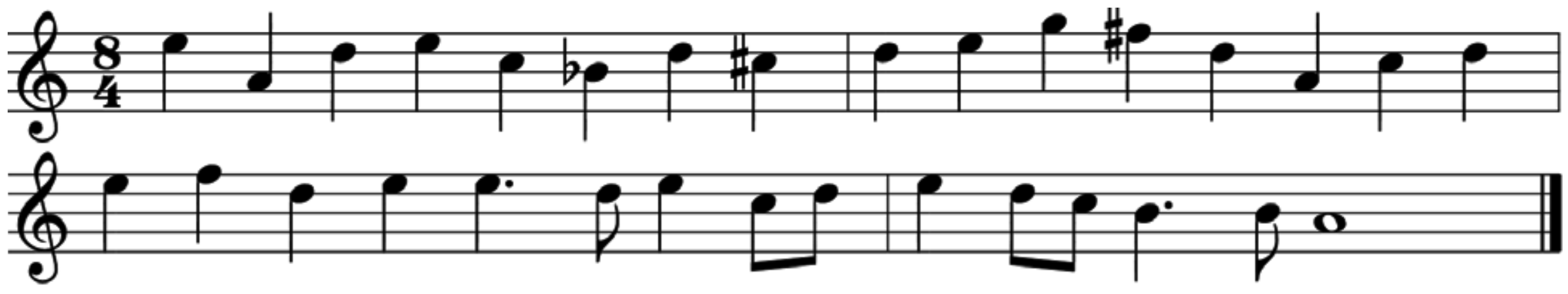}
    \caption{A training sequence.}
    \label{fig:trainingSequence}
\end{figure}

\begin{figure}[htp!]\centering
    \includegraphics[width=.9\textwidth]{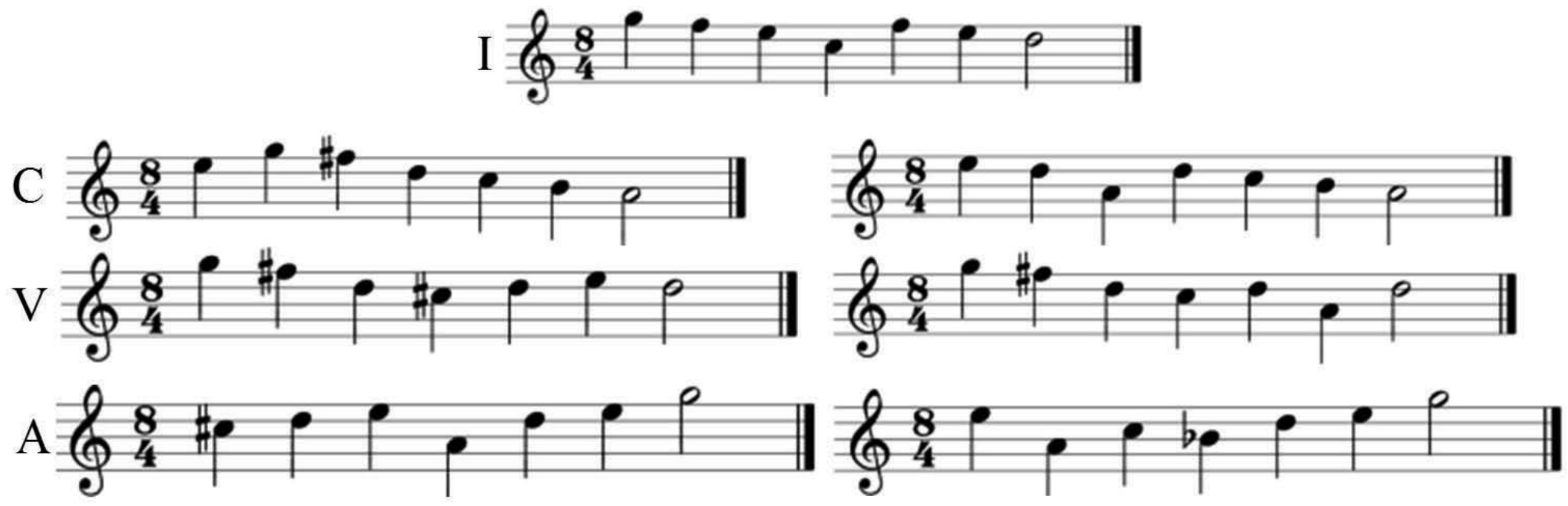}
    \caption{Different continuation types generated from the same training sequence: continuations (C), variations (V) and answers (A), generated from the input melody (I).}
    \label{fig:continuationTypes}
\end{figure}

Figure~\ref{fig:trainingSequence} shows a melody from which we build a Markov model $M$. Figure~\ref{fig:continuationTypes} shows examples of continuations, variations and answers, built from $M$ and the constraints corresponding to each melody type, from an input (I). It is clear that these melodies belong to their respective musical categories: Continuations end naturally, with the same 3 notes as the input (a consequence of the constraint on the last note); variations sound similar to the input; and answers sound as responses to the input. This shows how simple unary constraints can produce a global effect on the structure of generated melodies.

An application of positional constraints to the generation of jazz virtuoso melodies~\cite{pachet:09a} was also designed, controlled by a gesture controller. Sessions were recorded with various musicians~\cite{VirtuosoDInverno,VirtuosoSony,VirtuosoGeber}.

An interesting application of positional constraints was lyric generation, satisfying rhymes and prosody constraints~\cite{barbieri:12a,redylan}, leading to the generation of lyrics from arbitrary prosody and rhyme templates.

In all cases, the imposition of these positional constraints, and their ``back propagation'' consequence on the generated sequence, produce pleasing musical effects, probably due to the implicit generation of meaningful ``preparations'' for specific, imposed events. Note that this work was extended to handle positional constraints for recurrent neural networks in~\cite{DBLP:journals/nca/HadjeresN20}, leading to similar effects.

\subsubsection {Harmonization}
\label{section:harmonization}
\begin{figure}[htp!]\centering
    \includegraphics[width=1\textwidth]{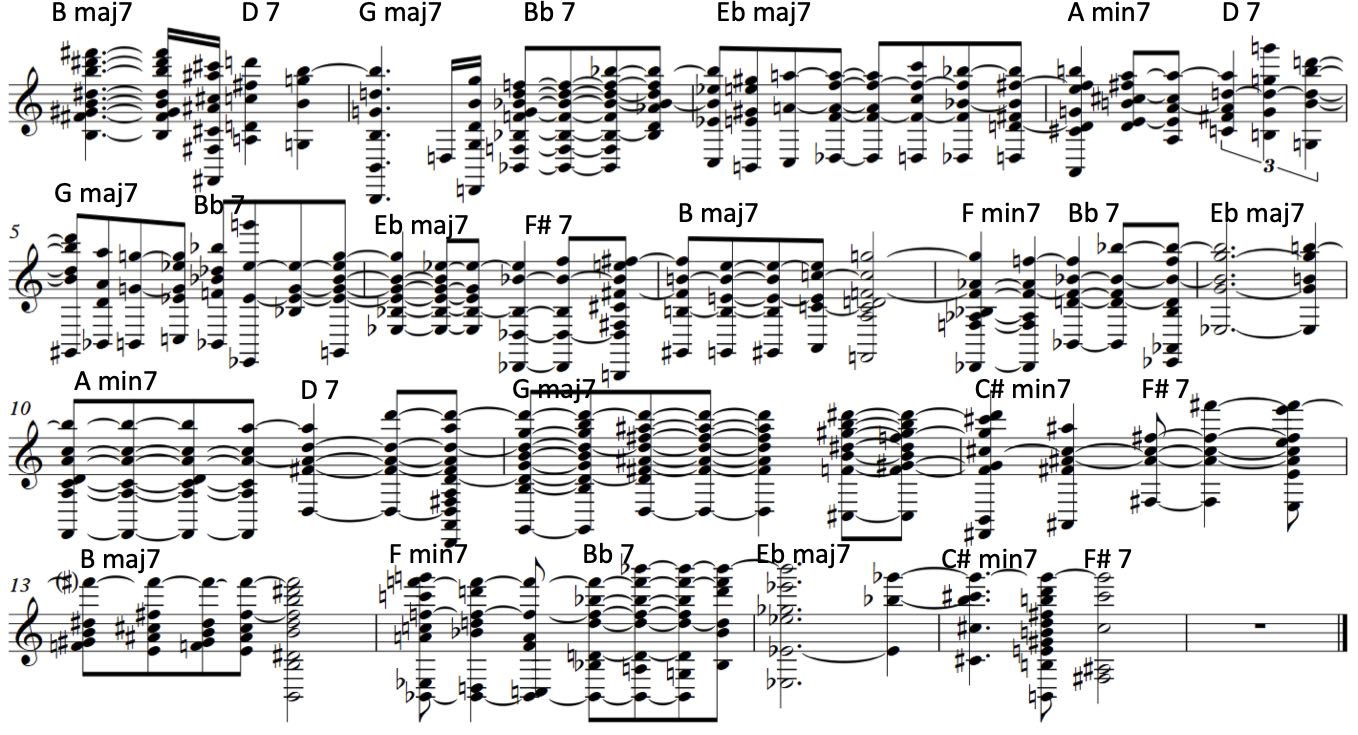}
    \caption{A harmonization of Coltrane's \emph{Giant Steps} in the style of Wagner. The melody is the soprano line, chord symbols are displayed, the rest is generated in the style of Wagner.}
    \label{fig:GiantStepsWagner}
\end{figure}

An interesting application of positional constraints was the harmonization system described in~\cite{pachet:14a}. In the system, positional constraints were used to enforce specific constraints on chords to be payed during the onset of melody notes. Passing chords were generated in between those notes to create interesting harmonic movements (called fioritures) that would fit with the imposed melody, but not necessarily in conformant ways. Some remarkable results were obtained, such as  a version of Coltrane's \emph{Giant Steps} \href{https://youtube.com/watch?v=Kq0ZwmOln7Y}{in the style of Wagner} (see~Figure~\ref{fig:GiantStepsWagner}).

An interesting harmonization of the title \emph{Come\c{c}ar de Nuovo} composed by Ivan Lins, in the style of Take 6 was produced, and shown to Ivan Lins himself, who reacted \href{https://www.youtube.com/watch?v=WBvX-03qP6A}{enthusiastically}.

These results obtained with positional constraints and Markov chains and their application paved the way for an extensive research roadmap, aimed at finding methods for controlling Markov chains with constraints of increasing complexity.

\subsection{Meter and all that jazz}
\label{section:meter}

The next problem to solve was meter: notes have durations and there are many reasons to constrain the sum of these durations in music. Meter, and more generally temporal properties, cannot be expressed easily by index positions in musical sequences, because events may have different positions. As a consequence, one cannot easily generate Markov sequences of events having variable durations, while imposing a fixed total duration for instance, which is problematic for music composition. Thanks to a theorem by Khovanskii in additive number theory~\cite{khovanskii:FAIA-1995}, we found a pseudo-polynomial solution for meter and Markov sequences. the \emph{Meter} constraint enables the generation of Markov sequences satisfying arbitrary constraints on partial sums of durations. This important result~\cite{DBLP:conf/aaai/RoyP13} was heavily used in our subsequent systems, notably for lead sheet generation ``in the style of'' with arbitrary constraints~\cite{pachet:14b}.

Another big step was to address a recurring problems of Markov chains: increasing the order leads to solutions which contain large copies of the corpus: how to limit this effect? MaxOrder was introduced in~\cite{DBLP:conf/aaai/PapadopoulosRP14} precisely to solve this problem, \ie use larger orders to increase imitation quality without generating plagiarism. The solution consists in reformulating the problem as an automaton, using the framework of regular constraints~\cite{DBLP:conf/cp/Pesant04}. We proposed a polynomial algorithm to build this automaton (MaxOrder imposes a set of forbidden substrings to the Markov sequences).

Now that we had a way to enforce basic constraints on meter and order, we investigated ways of making sequences more human. A beautiful result, in our opinion, was to revisit the classical result of Voss \& Clarke concerning $1/f$ distribution in music~\cite{VossClarkeNature,doi:10.1121/1.381721}. We looked for a constraint that biases a sequence so that its spectrum is in $1/f$. We showed that the stochastic, dice-based algorithm proposed by Voss can be expressed as a tree of ternary sum constraints, leading to an efficient implementation~\cite{DBLP:conf/ijcai/PachetRPS15}. For the first time, one could generate meter sequences in $1/f$. The examples in the original Voss paper did not have bars, understandably: now we could add them!

Paradoxically, the only negative result we obtained was to show that enforcing binary equalities within Markov chains was \#P-complete in the general case (as well as grammar constraints more generally)~\cite{DBLP:journals/corr/abs-1711-10436}. This is a counter intuitive result as this constraint seemed \emph{a priori} the simplest to enforce.

\subsection{Sampling methods}
The next class of problems we addressed was sampling. How to get not only all the solutions of a constraint problem, but a distribution of typical sequences? 
Works on sampling led to a remarkable result: all regular constraints (as introduced by Pesant in~\cite{DBLP:conf/cp/Pesant04}) added to Markov constraints can be sampled in polynomial time~\cite{DBLP:conf/cp/PapadopoulosPRS15}. 
We later realised that our positional constraint algorithm was equivalent to belief propagation~\cite{DBLP:conf/aaai/Pearl82}, and that meter, as well as \emph{MaxOrder}~\cite{papadopoulos2016generating} were regular. This led to a novel, faster and clearer implementation of metrical Markov sequences~\cite{DBLP:conf/cp/PapadopoulosPRS15}.

\begin{figure}[htp!]\centering
    \includegraphics[width=.9\textwidth]{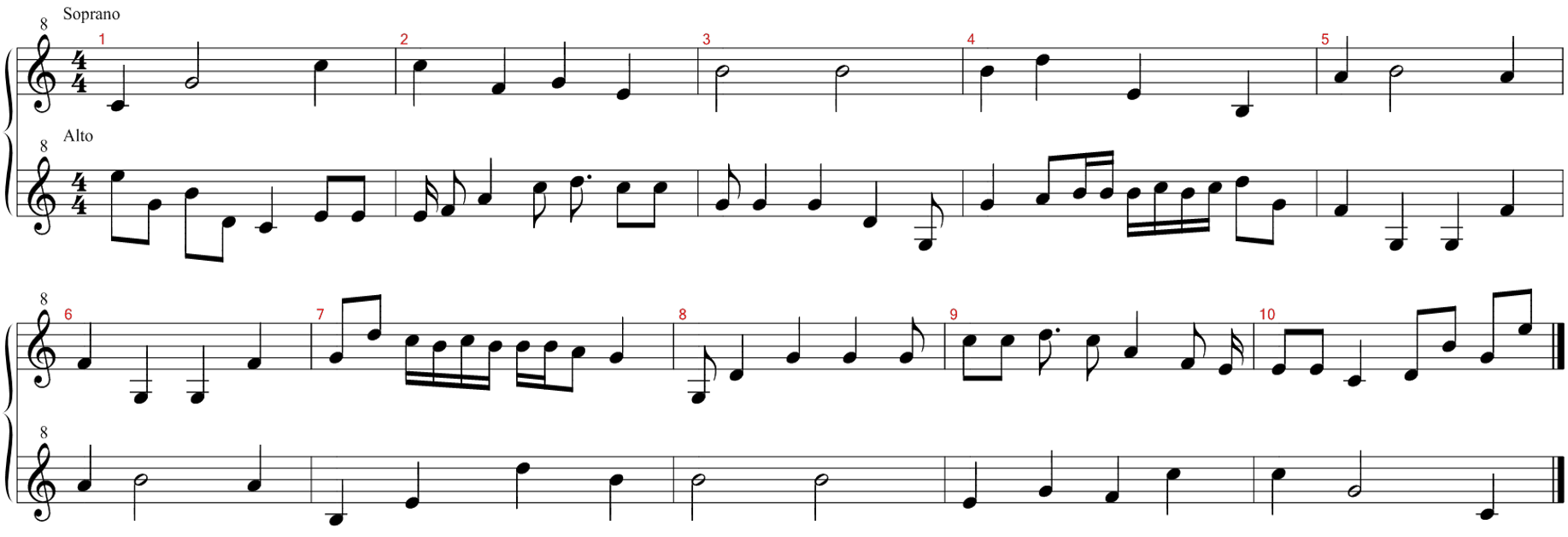}
    \caption{A 10-bar cancrizan canon generated by our palindrome Markov generator.}
    \label{fig:cancrizan-10bars}
\end{figure}

Some interesting extensions to more sophisticated constraints were studied, such as \emph{AllenMeter}~\cite{DBLP:conf/cp/RoyPRPPM16}. \emph{AllenMeter} allows users to express contraints using temporal locations (instead of events) involving all the Allen relations~\cite{DBLP:journals/cacm/Allen83}. This constraint was used to generate  polyphonic music enforcing complex synchronization properties (see section~\ref{section:Music}). Palindromes were also studied (\ie sequences that  read the same  forward and backward) and a beautiful graph-based solution was found to generate all palindromic sequences for 1st order Markov chains~\cite{DBLP:conf/ijcai/PapadopoulosRRP15}. A fascinating application to the generation of cancrizan canons was experimented by Pierre Roy with \href{https://www.francoispachet.fr/wp-content/uploads/2020/05/cancrizan2.mp3}{promising results} (see Figure~\ref{fig:cancrizan-10bars}).

Another interesting development addressed the issue of generating meaningful variations of musical sequences. This was performed by representing musical distance as biases of the local fields in the belief propagation algorithm~\cite{roy2017sampling,DBLP:conf/ismir/PachetPR17}. This type of issues is now addressed typically with variational auto-encoders~\cite{briot2019deep} with similar results.

These results somehow closed the chapter \emph{Markov + X} global constraints, since most interesting constraints in music and text can be expressed as a regular constraints.  This line of works generated many theoretical and algorithmic results~\cite{PachetEtAl}, as well as fruitful collaborations with other CP researchers. Jean-Charles R\'egin and Guillaume Perez, in particular, reformulated a number of our algorithms in the framework of Multi-valued Decision Diagrams (MDD), yielding substantial gains in efficiency~\cite{DBLP:conf/cp/PerezR17,DBLP:conf/cp/PerezRG18}.

\section{Beyond Markov Models}
\label{section:beyondMarkov}

Markov models (and their variable order species) having been thoroughly investigated, we turned to more powerful models with the same goal in mind: finding efficient ways of controlling them. We  explored the use of the \emph{maximum entropy} principle~\cite{jaynes57} for music modelling. This model is based on the representation of binary relationships between possibly not contiguous events, thereby preventing issues related to parameter explosion inherent to higher-order Markov models. Departing from a pure filtering approach, parameter estimation is performed using high-dimension gradient search, and generation using a Metropolis algorithm.  This model gave interesting results for monophonic sequences~\cite{sakellariou2017maximum} and some extensions were studied to polyphonic ones as well~\cite{DBLP:journals/corr/HadjeresSP16}. An application to modelling expressiveness in monophonic melodies was conducted in~\cite{DBLP:journals/corr/MoulierasP16} with promising results.

Other aspects of style capture and generation were considered in Flow Machines, such as the generation of audio accompaniments ``in the style of''. A dynamic programming approach was developed in conjunction with a smart ``audio gluing'' mechanism to preserve \emph{groove}, \ie small deviations in the onset of events that characterize the style~\cite{DBLP:conf/ijcai/RamonaCP15}. A convincing example can be heard in the Bossa nova orchestration of \emph{Ode to Joy}~\cite{DBLP:journals/tist/Pachet17}. This result triggered a fruitful collaboration with Brazilian colleagues to capture Brazilian guitar styles (the Brazyle project~\cite{BrazyleProject}).

\section {Flow Composer: the first AI assisted lead sheet composition tool}
These techniques were used to develop a lead sheet generator called Flow Composer (see~\cite{CommentsOnFlowComposer} for a retrospective analysis of the development of this project).
 The lead sheet generator was trained using a unique database of lead sheets developed for this occasion, LSDB~\cite{lsdb:ismir:2013}.
In order to use these generators, we designed the interface Flow Composer~\cite{papadopoulos:flow:composer:cp:2016}. The basic idea was to let users enter arbitrary chunks of songs (melody or chords) and let the generator fill in the blanks, a process referred to now as \emph{inpainting}~\cite{FlowComposerVideo}. This process took several iterations, ranging from the development of a javascript library for music web editing~\cite{MartinTenor} to studies about the impact of feedback on composition with these tools~\cite{MartinFeedback}.

Thanks to this interface, Flow Composer was intensively used by musicians, in particular by \href{https://open.spotify.com/artist/4aGSoPUP1v4qh7RfBlvgbR/about}{SKYGGE} (the artist name Benoit Carr\'e uses for all musical productions done with AI) to compose and produce what turned out to be the first mainstream music album composed with AI: \href{https://www.helloworldalbum.net}{Hello World}.

\section{Significant Music Productions}
\label{section:Music}
In this section, we review some of the most significant music produced with the tools developed in the Flow Machines project. Departing from most research in computer music, we stressed from the beginning the importance of working with musicians, in order to avoid the ``demo effect'', whereby a feature is demonstrated in a spectacular but artificial way, regardless of the actual possibility by a real musician to use it to make real music to a real audience.
It is the opinion of the authors that some of the music described above stands out, but we invite the reader to listen to these examples to form his/her judgement.

\begin{enumerate}

\item The \emph{Boulez Blues}

The \emph{Boulez Blues} (see section~\ref{section:MarkovPlusXRoadmap}) is a Blues chord sequence in the style of Charlie Parker (generated from a 1st order Markov model). The sequence contains only major and minor chords, in all 12 keys (\ie 24 chords in total, 2 per bar) and additionally satisfies an \emph{AllDifferent} constraint, \ie all chords are different. The sequence was generated with a preliminary version of Markov constraints, that enabled the computation of the most probable sequence (using branch \& bound). In that sense, we can say that the \emph{Boulez Blues} is the most Parker-like Blues for which all 24 chords are different (check this \href{https://www.francoispachet.fr/wp-content/uploads/2020/04/boulez_blues.mp3}{rendering with jazz musicians}). This remarkable Blues chord progression can be considered as an original \textcolor{red}{stylistic singularity} (of the style of Charlie Parker, using a specific constraint which clearly goes in the way of the style).

\item Two harmonizations with Flow Composer

Some harmonizations produced with Flow Composer (see Section~\ref{section:harmonization}) stand out. We stress here the musical qualities of the harmonization of \emph{Giant Steps} \href{https://youtube.com/watch?v=Kq0ZwmOln7Y}{in the style of Wagner}, and the harmonization of \emph{Come\c{c}ar de Nuovo} (Ivan Lins) \href{https://www.youtube.com/watch?v=WBvX-03qP6A}{in the style of Take 6}. These pieces can be seen as particular instances of \textcolor{red}{style transfer} (here the style of orchestration is transferred from one piece to another).

\item Orchestrations of \emph{Ode to Joy}

During the project, at the request of the ERC for the celebration of the 5000th grantee, we produced seven orchestrations of Beethoven's \emph{Ode to Joy} (the anthem of the European Union). These orchestrations were produced with various techniques and in different styles~\cite{DBLP:journals/tist/Pachet17} and can be \href{https://www.youtube.com/watch?v=buXqNqBFd6E}{heard online}.

The most notable orchestrations are:

\begin{enumerate}
\item Multi-track audio generation
A particularly interesting application of the work on \emph{AllenMeter} by Marco Marchini (see section~{\ref{section:meter}}) is the generation of multi-track audio, involving temporal synchronization constraints between the different tracks (bass, drum, chords). Figure~\ref{fig:multiTrackAudio} shows an excerpt of \emph{Ode to Joy} in the style of \emph{Prayer in C} (by Lilly Wood and the Prick). 

\begin{figure}[htp!]\centering
    \includegraphics[width=.9\textwidth]{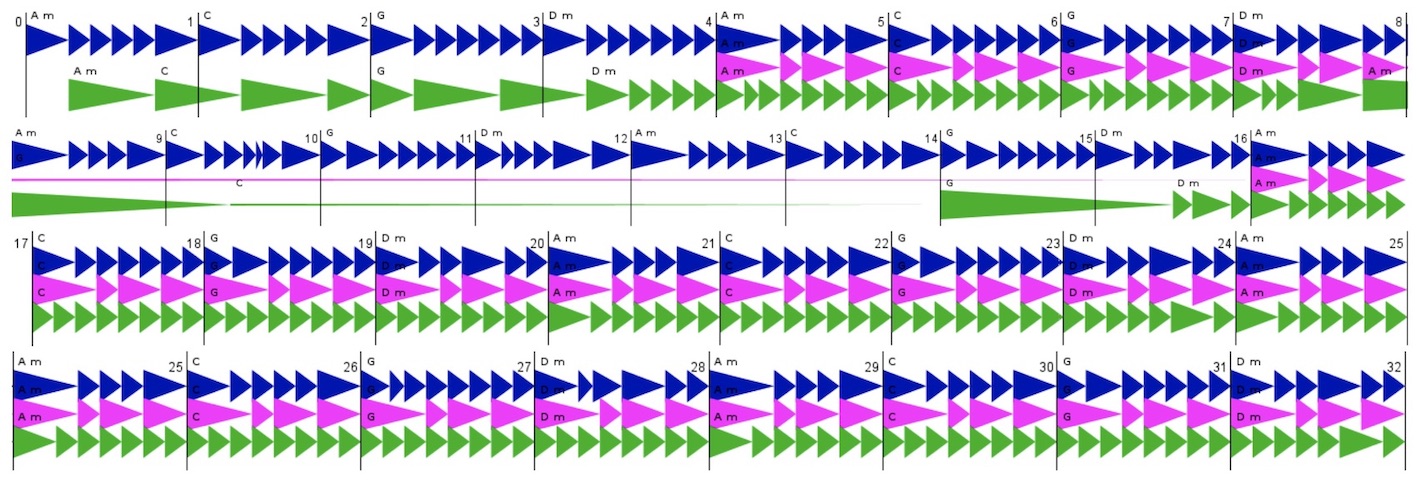}
    \caption{A multi-track audio piece generated with Allen Meter: A graphical representation of the guitar (top), bass (center), and drum (bottom)
tracks of \emph{Prayer in C.} Each track contains 32 bars and each triangle represents a chunk. Vertical lines indicate bar separations.}
    \label{fig:multiTrackAudio}
\end{figure}

\item Bossa Nova orchestrations with concatenative synthesis

A Bossa nova orchestration was generated for which the guitar accompaniment was produced using concatenative synthesis, applied to a corpus of guitar recordings (see Section~\ref{section:beyondMarkov}). The techniques used are described in~\cite{DBLP:conf/ijcai/RamonaCP15}, and this Bossa nova example nicely emphasizes  how the groove of the original guitar recording (by the first author of this paper) was preserved.

\item Bach like chorale using Max Entropy

An orchestration in the style of Bach was generated, using the maximum entropy principle (see Section~\ref{section:beyondMarkov}), which had shown great theoretical as well as musical results on monophonic music modelling~\cite{sakellariou2017maximum}. This anachronic yet convincing orchestration paved the way for the DeepBach system, a more complete orchestration system in the style of Bach based on an LSTM architecture~\cite{DBLP:conf/icml/HadjeresPN17}.

\end{enumerate}

\item \emph{Beyond The Fence}

\emph{Beyond The Fence} is a unique musical, commissioned by Wingspan Production. The idea was to produce a full musical show using various \emph{computational creativity} techniques, from the pitch of the musical to the songs, including music and lyrics. The musical was produced in 2015 and was eventually staged in the Arts Theatre in London’s West End during February and March of 2016.
The complex \emph{making of} of the musical gave birth to two 1-hour documentary films, which were aired on Sky Arts under the title \emph{Computer Says Show}.

Some of the songs were composed with a preliminary version of Flow Composer, under the control of composers Benjamin Till and Nathan Taylor.
In particular, the song \emph{Scratch That Itch} (see Figure~\ref{fig:scratchThatItch}) received good criticism.
The song was produced by training Flow Composer on a corpus of musicals, mostly broadway shows, in an attempt to replicate that style.

\begin{figure}[htp!]\centering
    \includegraphics[width=.9\textwidth]{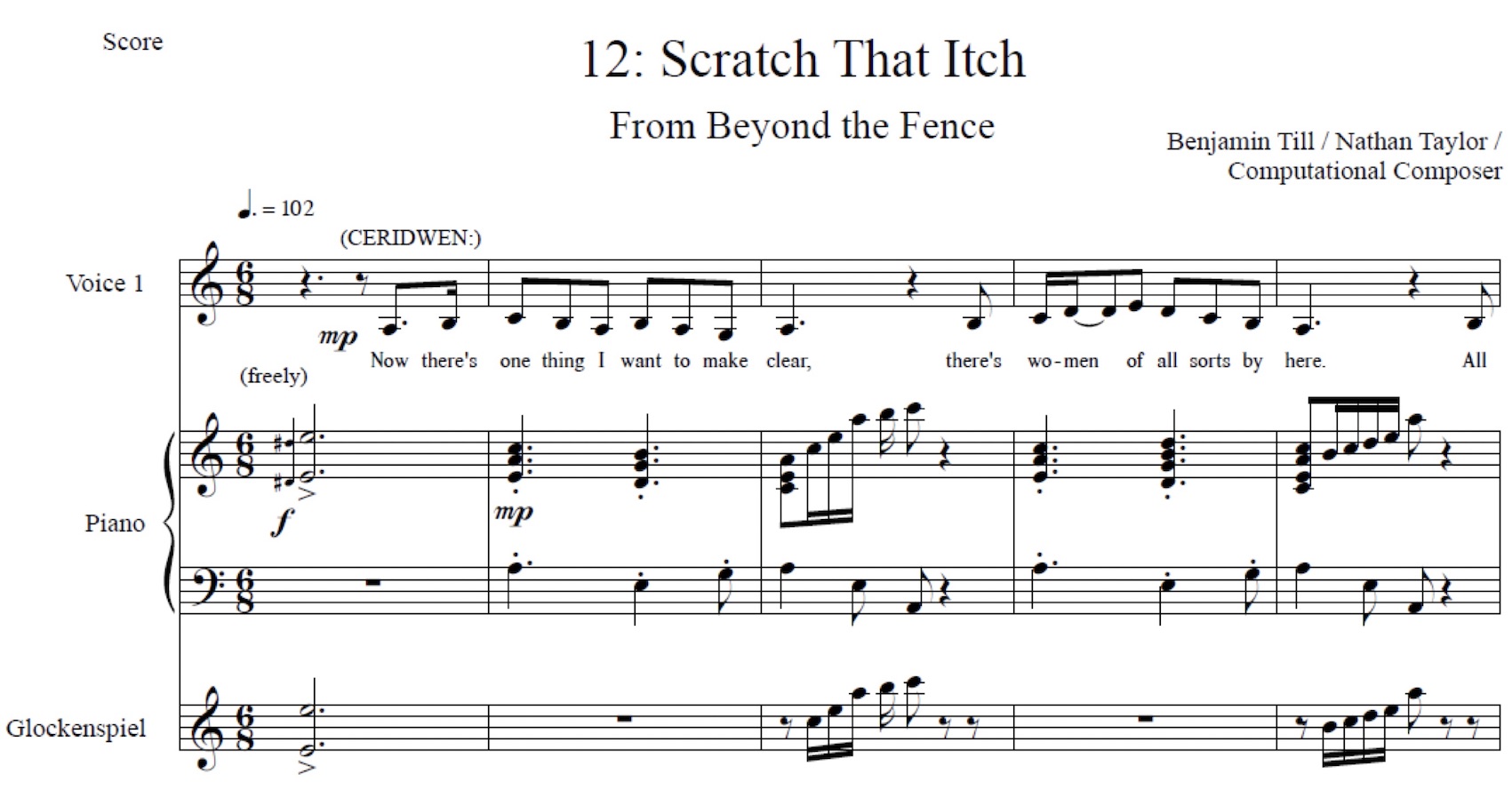}
    \caption{An excerpt of \emph{Scratch That Itch}, a song composed with Flow Composer for the musical \emph{Beyond The Fence}.}
    \label{fig:scratchThatItch}
\end{figure}

An analysis of the production and reception of the musical~\cite{colton:16} showed that the overall reception was good, though difficult to evaluate, because of the large scope of the project. Concerning the songs made with Flow Composer, a critic wrote:

\begin{quote}
I particularly enjoyed Scratch That Itch, \dots, which \textcolor{red}{reminded} me of a Gilbert \& Sullivan number whilst other songs have elements of Les Miserables. Caroline Hanks-Farmer, carns, TheatrePassion.com
\end{quote}

This critic embodies the essence of the reception in our view: songs created did bear some analogy with the corpus, and created a feeling of \textcolor{red}{reminiscence}, at least to some listeners.

\item Catchy tune snippets with Flow Composer

Before Flow Composer was used on a large scale for the \emph{Hello World} album (see below), the system was used to compose fragments of songs in various styles. Some of these fragments are still today worth listening, bearing a catchiness that is absent from most of the generated music produced 4 years later. These three examples can be seen as \textcolor{red}{stylistic explorations} of various composers.

\begin{enumerate}
\item in the style of Miles Davis

This \href{https://www.francoispachet.fr/wp-content/uploads/2020/04/MilesGetLucky.mp3}{song snippet} was generated by training Flow Composer on a set of about 10 songs composed by Miles Davis (see Figure~\ref{fig:MilesStyle}). These 8 bars, played in loop, produce an engaging groove though with a rather unconventional yet consistent melody.

\begin{figure}[htp!]\centering
    \includegraphics[width=.9\textwidth]{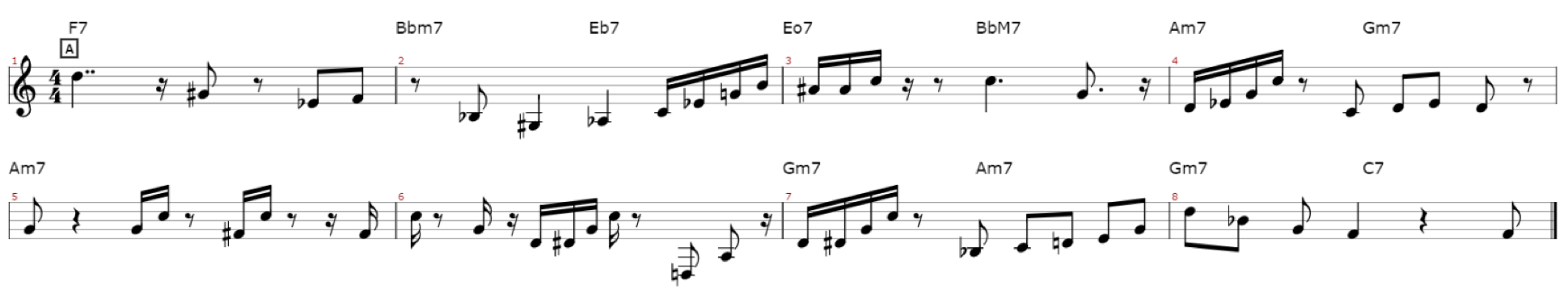}
    \caption{A song snippet composed with Flow Composer in the style of Miles Davis~\cite{MilesDavisSnippet}.}
    \label{fig:MilesStyle}
\end{figure}

\item From \emph{Solar} to \emph{Lunar}

This \href{https://www.francoispachet.fr/wp-content/uploads/2020/04/Miles-Davis-Mix_DEF.mp3}{short song} was also generated by training Flow Composer on a set of 10 songs composed by Miles Davis (see Figure~\ref{fig:MilesStyle}). Additionally, the structure of the song \emph{Solar} was reused, an instance of \textcolor{red}{templagiarism}, hence its title. This song was deemed sufficiently interesting to be \href{https://www.francoispachet.fr/wp-content/uploads/2020/04/LunarPerformance.mp3}{performed live in London} (Mark d'Inverno quartet, see Figure~\ref{fig:VortexConcert}) during an ``AI concert''.
Listening to the the tune performed by Mark d'Inverno quartet, journalist James Vincent wrote~\cite{TheVergeJamesVincent}:
\begin{quote}
The AI’s contribution was just a lead sheet --- a single piece of paper with a melody line and accompanying chords --- but in the hands of d'Inverno and his bandmates, it swung. They started off running through the jaunty main theme, before devolving into a series of solos that d’Inverno later informed me were \emph{all human} (meaning, all improvised).
\end{quote}
 
\begin{figure}[htp!]\centering
    \includegraphics[width=1\textwidth]{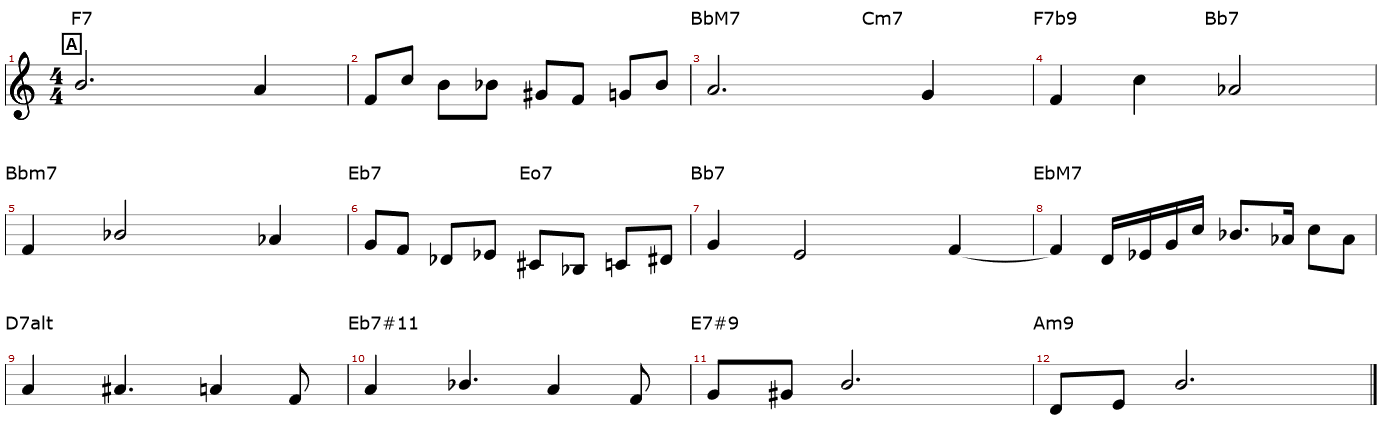}
    \caption{\emph{Lunar}, a song composed with Flow Composer in the style of Miles Davis, with the structure of Solar~\cite{LunarRendering,LunarPerformance}.}
    \label{fig:Lunar}
\end{figure}

\begin{figure}[htp!]\centering
    \includegraphics[width=.9\textwidth]{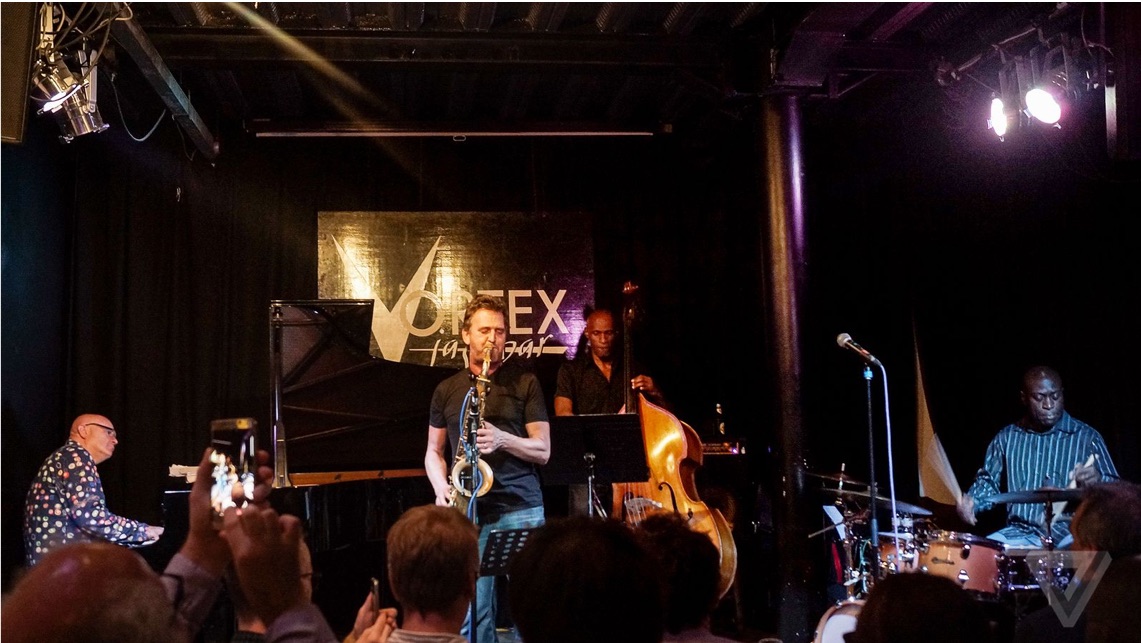}
    \caption{The concert of the Mark d'Inverno quartet playing \emph{Lunar} at the Vortex Club in London, October 2016.}
    \label{fig:VortexConcert}
\end{figure}

\item in the style of Bill Evans

This \href{https://www.francoispachet.fr/wp-content/uploads/2020/04/BillEvansStyle.mp3}{short song} was generated by training Flow Composer on a set of 10 songs composed by Bill Evans (see Figure~\ref{fig:BillEvansStyle}). The melody gracefully navigates through the chord progression, sometimes in an unconventional but clever manner (bar 3). Thanks to unary constraints, the tune nicely transitions from the end to the beginning. Other songs in the style of Bill Evans were generated on the fly and performed live at the Ga\^it\'e Lyrique concert (see below).

\begin{figure}[htp!]\centering
    \includegraphics[width=.9\textwidth]{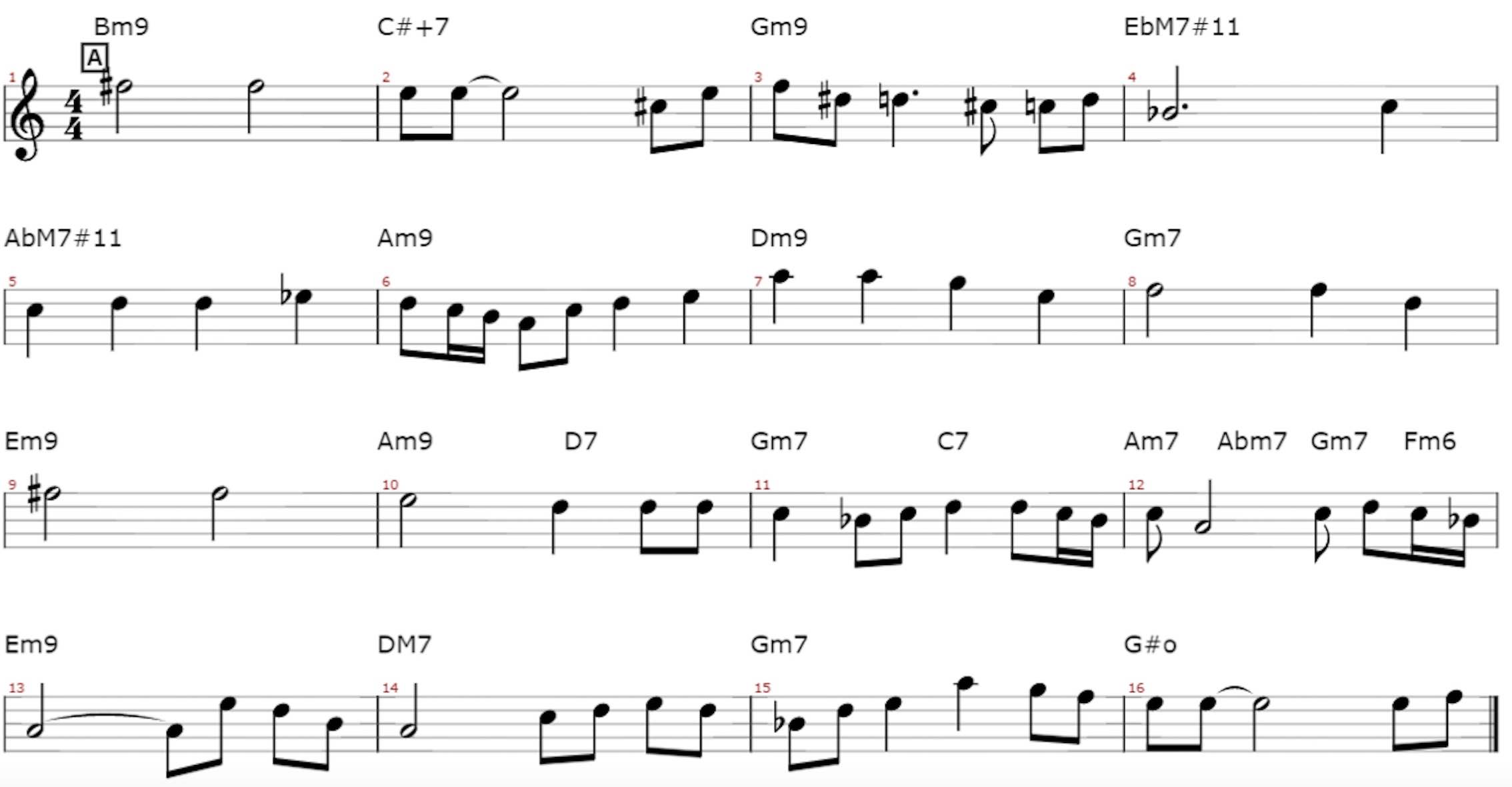}
    \caption{A song snippet composed with Flow Composer in the style of Bill Evans. Available at~\cite{BillEvansStyle}.}
    \label{fig:BillEvansStyle}
\end{figure}

\end{enumerate}

\item \emph{Daddy's car}

The song \emph{Daddy's car} was generated with Flow Composer, using a corpus of 45 Beatles song (from the latest period, considered usually as the richest and most singular in the recording history of the Beatles). Flow Composer was used to generate the lead sheet, while the lyrics and most of the orchestration were done manually, by SKYGGE.
The song was \href{https://www.youtube.com/watch?v=LSHZ_b05W7o}{released in septembre 2016 on YouTube}~\cite{DaddysCarVideo} and received substantial media attention\footnote{About 2.5 million views as of April 2020}.

An interesting aspect of this song is how the system identified and reproduced a typical harmonic pattern found in many Beatles song. This pattern consists in a chromatic descending bass and can be found for instance in the introduction and in the middle of the song \emph{Michelle} (which was part of the training set) (see Figure~\ref{fig:MichelleExtract}). This pattern was reused and instantiated in a novel context in the middle of \emph{Daddy's car} (see Figure~\ref{fig:DaddysCarExtract}).
For that reason, the song can be considered a \textcolor{red}{pastiche}: it reuses explicitly typical patterns of the style, without any attempt at introducing  novelty.

\begin{figure}[htp!]\centering
    \includegraphics[width=.9\textwidth]{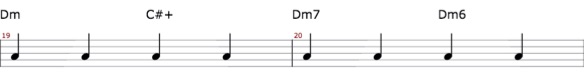}
    \caption{The introduction of \emph{Michelle} uses a typical harmonic pattern of the Beatles (here in D minor). Note that we show here the Real Book spelling, with a C\#+ chord, which is equivalent to a DmM7.}
    \label{fig:MichelleExtract}
\end{figure}

\begin{figure}[htp!]\centering
    \includegraphics[width=.9\textwidth]{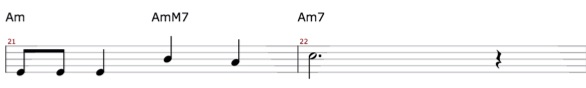}
    \caption{The song \emph{Daddy's car} reuses the harmonic pattern of \emph{Michelle} in a different context (here in A minor).}
    \label{fig:DaddysCarExtract}
\end{figure}


\item Busy P remix

Artist Busy P (real name \href{https://en.wikipedia.org/wiki/Pedro_Winter}{Pedro Winter}, French DJ, producer and ex-manager of Daft Punk), launched title \href{https://open.spotify.com/track/7J02VXGmyD2OuGPLTnuONu?si=QKqpmXe_SACtpGsI81Okbw}{\emph{Genie}} in 2016~\cite{GenieBusyP}. Several remixes of the title were produced, including \href{https://www.youtube.com/watch?v=Pf_1_wDeoSY}{one made with Flow Machines}~\cite{GenieRemix}. In this remix, a variation of the chord sequence was generated with Flow Composer by SKYGGE, and then rendered using concatenative synthesis~\cite{DBLP:conf/ijcai/RamonaCP15}. The remix features an interesting and surprising slowdown, due to an artefact in the generation algorithm, that was considered creative and kept by the remixer.

\item \emph{Move on}

Unesco commissioned a song to be composed with Flow Machines, using material representative of 19 so-called ``creative cities''.
This song, entitled \href{https://soundcloud.com/user-76901948/sets/la-playlist-des-villes-creatives-de-lunesco}{\emph{Move On}}, was composed in 2017 and distributed as a physical 45 RPM vinyl~\cite{MoveOn2017}. 20 scores of songs were gathered and used as a training set. The composition and realization was performed by SKYGGE, Arthur Philippot and Pierre Jouan, with the band \href{http://lacatastrophe.fr/}{Catastrophe}.
The song integrates a large number of musical elements, sounds and voices but manages to produce a coherent and addictive picture, thanks to the repeated use of smart transformations of the sources to fit the same chord progression throughout.

\item Concert at Ga\^it\'e Lyrique

A unique concert involving several artists who composed songs with Flow Machines was performed at Ga\^it\'e Lyrique in Paris, in October 2016. This concert was probably the first ever concert showcasing pop tunes (as well as jazz) composed (in real time for some of them) with AI. 

The \href{{https://www.youtube.com/watch?v=bptKZ2ACZfQ}}{concert}~\cite{GaiteLyriqueConcert2016} involved the following artists and songs:

\begin{enumerate}
\item SKYGGE (\emph{Daddy's Car})

SKYGGE \href{{https://www.youtube.com/watch?v=cTP0Sr_ehmY}}{performed \emph{Daddy's car}} as well as several titles~\cite{DaddysCarConcert2016} which appeared later in the \emph{Hello World} album (see below for descriptions).

\item Camille Bertault

Jazz singer \href{https://www.camillebertault.fr/}{Camille Bertault} (who has, since then, become a successful jazz interpret) sung \href{https://www.youtube.com/watch?v=1YfKbLcjwUM}{songs composed in real-time} by Flow Composer, in the style of Bill Evans~\cite{BertaultConcert2016}. The voice of Bertault fits particularly well with the Evanssian harmonies and the simple but efficient voicings played by the pianist (Fady Farah).

\item O (\emph{Azerty})

Artist Olivier Marguerit (aka ``O'') composed a \href{https://www.youtube.com/watch?v=v69KcJBrpuo}{beautifully original song} entitled \emph{Azerty}~\cite{AzertyConcert2016}. Verses were generated by Flow Composer trained on a mix of \emph{God Only Knows} by The Beach Boys, \emph{Sea Song} by Robert Wyatt and \emph{The Sicilian Clan} by Ennio Morricone. 
This song features interesting, and in some sense typical, melodic patterns created with Flow Composer that sound momentarily unconventional but eventually seem to resolve and make sense, pushing the listener to keep his attention on the generously varied melodic line. The song also features a 4 bar melody woven onto an unconventional harmony that deserves attention (see Figure~\ref{fig:AzertyMelody}).

\begin{figure}[htp!]\centering
    \includegraphics[width=1\textwidth]{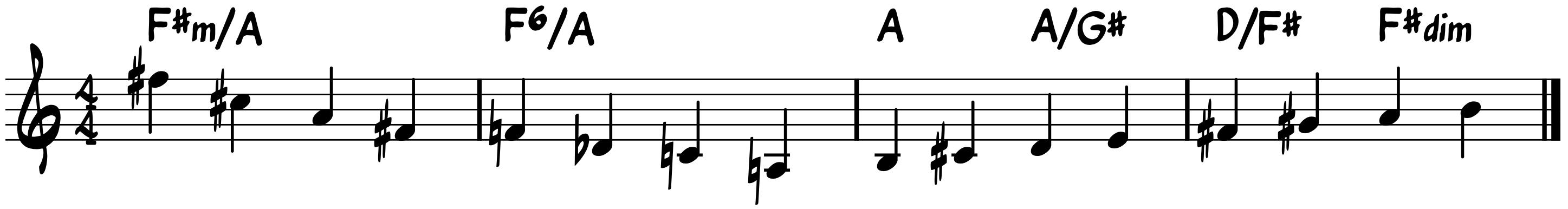}
    \caption{A haunting 4 bar melody/harmony combination at the heart of the song Azerty.}
    \label{fig:AzertyMelody}
\end{figure}

\item Fran\c{c}ois Pachet (Jazz improvisation with Reflexive Looper)

Pachet performed a \href{https://www.youtube.com/watch?v=8YzPaCzDDzg}{AI-assisted guitar improvization} on the jazz standard \emph{All the things you are} using Reflexive Looper~\cite{PachetConcert2016}, a system that learns in real time how to generate bass and chord accompaniments~\cite{DBLP:conf/chi/PachetRMd13,Marchini2017RethinkingRL} (see section~\ref{section:Unfinished}).

\item Kumisolo (\emph{Kageru San})

Japanese artist Kumisolo interpreted the song \href{https://www.youtube.com/watch?v=sCmp3Xp_PG4}{\emph{Kageru San}}, composed in the JPop style by SKYGGE with Flow Composer~\cite{KumisoloConcert2016}.
Flow Composer was trained with songs from the album \emph{Rubber Soul} by The Beatles.  The audio renderings were generated from excerpts of \emph{Everybody's Talking at Me} by Harry Wilson and \emph{White Eagle} by Tangerine. The song, while firmly anchored in the JPop style, has a memorizable, catchy chorus.

\item Lescop (\emph{A mon sujet})

The song \href{https://www.youtube.com/watch?v=9uTIyJ8vDRM}{\emph{A mon sujet}}~\cite{LescopConcert2016} was composed by Lescop with Flow Composer trained on songs \emph{Yaton} by Beak, \emph{Vitamin C} by Can, \emph{Tresor} by Flavien Berger, \emph{Supernature} by Cerrone and \emph{Rain} by Tim Paris. Although the resulting melodic and harmonic lines are monotonous compared to other songs, the voice of Lescop mixed with the piano and bass ostinato create a hauntingly obsessive and attaching musical atmosphere.

\item ALB (\emph{Deedadooda})

The song \href{https://www.youtube.com/watch?v=u7iG4SKbIbY&list=PLvoqwxjRRNfmn2A6e9-gdWo9ORhZfLFmO&index=4}{\emph{Deedadooda}}~\cite{ALBConcert2016} by emerging artist ALB was composed with a training set of 16 songs by ALB himself. It features a dialog between ALB and a virtual composer, cleverly integrated in the song itself.

\item Housse de Racket (\emph{Futura})

The song \href{https://www.youtube.com/watch?v=BU9g0X_H8zo}{Futura} was composed from a training set of six songs by ``Housse de Racket'' themselves~\cite{HousseDeRacketConcert2016}. The rendering is generated from the introduction of \emph{Love's in Need of Love Today} by Steve Wonder, which creates an enigmatic and original Pop atmosphere on an unusually slow tempo.

\end{enumerate}

\item \emph{Hello World}

The most sophisticated music production of the project was the album \href{https://www.helloworldalbum.net}{\emph{Hello World}}~\cite{HelloWorldAlbumSite}. \emph{Hello World} was the first mainstream music album featuring only AI-assisted compositions and orchestrations. This album features 15 songs composed with various artists who used Flow Machines at its latest stage. Composition sessions were held at Sony CSL, conducted by SKYGGE, and involved artists in several genres (Pop, Jazz, Electronic Music). \emph{Hello World} received considerable \href{https://www.helloworldalbum.net/press}{media attention}~\cite{HelloWorldMedia} and reached remarkable audience stream counts (about 12 million streams). 
Each song has its own motivation and \emph{story}, which we summarize as follows\footnote{songs can be heard at \url{https://skyggewithai.bandcamp.com/album/hello-world} as well as on all streaming platforms. Interesting \href{https://skyggewithai.bandcamp.com/album/hello-world-demos} {demo material} has been released publicly as well.}:

\begin{enumerate}[label={(\arabic*)}]

\item \href{https://skyggewithai.bandcamp.com/track/ballad-of-the-shadow}{\emph{Ballad of the Shadow}}

\emph{Ballad of the Shadow} is the first song written for this album. It was composed by SKYGGE with Flow Machines, inspired by jazz standards, with the goal of making a vaporwave cowboy song. The idea of shadows singing was developed into a melody with a happy mood. It resembles a cartoon-like tune when it is played at 120 bpm, but becomes melancholic when played slower, especially with ambient textures. Ash Workman and Michael Lovett detuned the drums and added some drive effects. The main theme (see Figure~\ref{fig:BalladOfTheShadow}) is, again, enigmatic, unconventional yet beautifully makes its way through an unusual chord progression.

\begin{figure}[htp!]\centering
    \includegraphics[width=.9\textwidth]{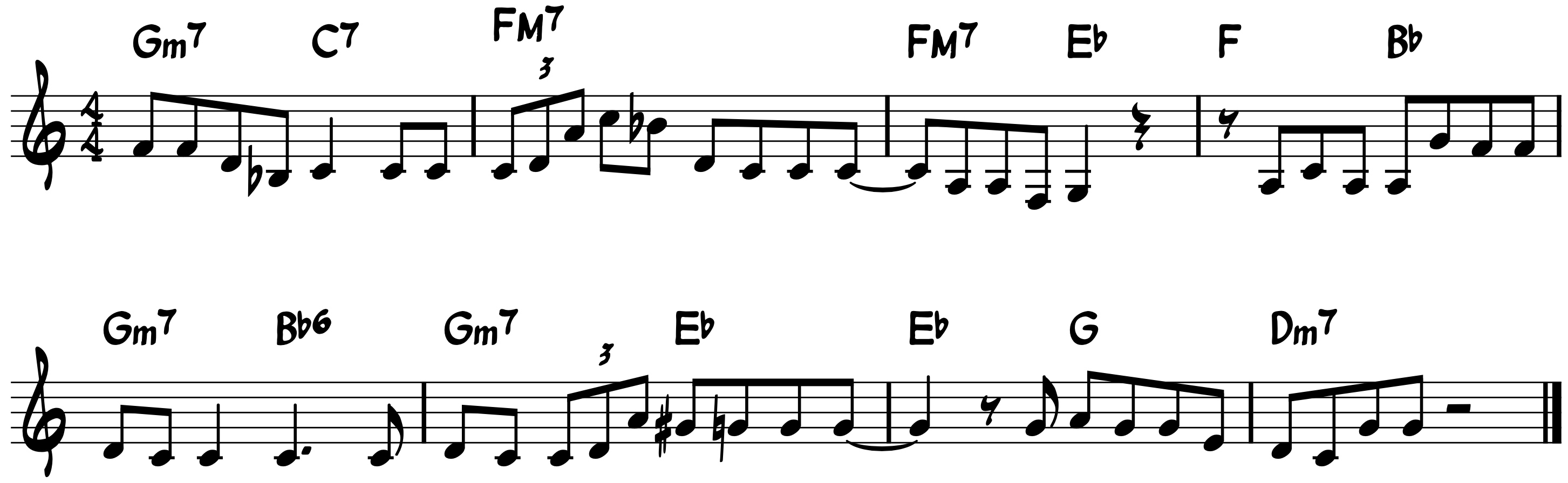}
    \caption{An enigmatic 8 bar theme at the heart of \emph{Ballad of the Shadow}.}
    \label{fig:BalladOfTheShadow}
\end{figure}

\item \href{https://skyggewithai.bandcamp.com/track/sensitive-feat-c-duncan}{\emph{Sensitive}} 

\emph{Sensitive} was composed by SKYGGE with Flow Machines, inspired by bossa novas. Lyrics and voice are from C Duncan.
The song was composed from a corpus of bossa novas of the 60s. There are some patterns in the song such as major-minor progressions, that bossa nova fans will recognize and like (see Figure~\ref{fig:Sensitive}). Harmonic changes are sometimes very audacious, but the melody always stays on tracks. SKYGGE generated a voice for the melodic line with random lyrics, like a mosaic of syllables extracted from the a cappella recording of a vocalist. However, once played with a piano, the music sounded like a powerful 70s ballad. SKYGGE liked the combination of the two styles, and recorded a string arrangement written by Marie-Jeanne Serrero and live drums. C Duncan was very enthusiastic when he listened to the song, he wrote lyrics from random material inspired by key words like \emph{wind}, \emph{deep} and \emph{feel}. 

\begin{figure}[htp!]\centering
    \includegraphics[width=1\textwidth]{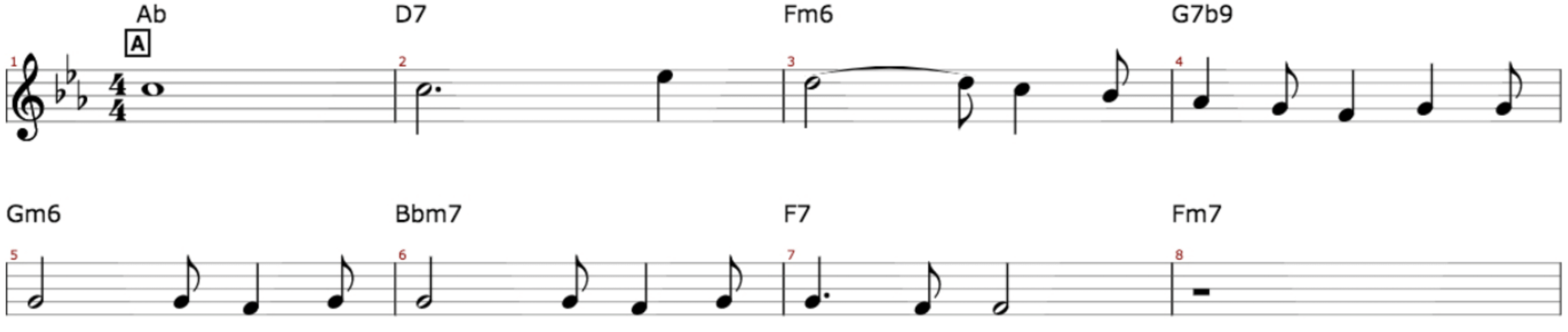}
    \caption{The first bars of \emph{Sensitive}, a song composed in the style of Bossa Novas, but rendered in a slow pop ballad.}
    \label{fig:Sensitive}
\end{figure}

\item \href{https://skyggewithai.bandcamp.com/track/one-note-samba-feat-generated-vocals-from-the-pirouettes}{\emph{One Note Samba}}

Fran\c{c}ois Pachet and SKYGGE share an endless admiration for Antonio Carlos Jobim, the great Brazilian composer, and they wanted to do a cover of his famous song \emph{One Note Samba} with Flow Machines. 

SKYGGE put together generated stems with drums, bass and pads and right away got awesome results: a singular and catchy tune with great harmonies and timbre. The resulting harmonies slightly differ from the original but do not betray the logic of the song. The chord progression brings a Jobimian touch to the melody through unexpected harmonic modulations.

A few days before, the French band \emph{The Pirouettes} had come to the studio and had uploaded two songs from their first album \emph{Carr\'ement, Carr\'ement}. Vocals were generated from these recordings for \emph{One Note Samba}, so fans of The Pirouettes may recognize words from their original songs.

\item \href{https://skyggewithai.bandcamp.com/track/magic-man}{\emph{Magic Man}}

SKYGGE fed Flow Machines with French pop songs from the 80s. The machine generated a simple melody, which sounded groovy when rendered it with a generated choir. The title comes from a phrase that comes back frequently in the choir: \emph{Magic Man}. It was a nice surprise that the machine came up with a shiny pop song title with such an electro-disco feel. Flow Machines generated guitars from an American folk stem, as well as other vocals on the verse, and SKYGGE sung over those voices to get a more complex vocal blend. He also asked the singer Mariama to sing along with the generated choir to reinforce the groove. The lyrics are a mashup from all the generated syllables. French electro band Napkey worked on the arrangement at the end of the production, and Michael Lovett added synthesizer arpeggios.

\item \href{https://skyggewithai.bandcamp.com/track/in-the-house-of-poetry-feat-kyrie-kristmanson}{\emph{In the House of Poetry}}

SKYGGE wanted to compose a song with the enchanting charm of ancient folk melodies. He fed Flow Machines with folk ballads and jazz tunes. As a result, the machine generated melodies with a chord progressions right in that mood, and a catchy and singular melodic movement (see Figure~\ref{fig:HouseOfPoetry1}). Once the verse was done, he fed Flow Machines with a jazzier style for the chorus part, in order to bring in rich harmonic modulations.

Flow Machines then proposed an unconventional and audacious harmonic modulation in the first bar of the chorus, with an ascending melody illuminating the song (see Figure~\ref{fig:HouseOfPoetry2}). SKYGGE followed up with a small variation by exploiting a 2-bar pattern generated by Flow Machines and asked the system again to generate a harmonic progression that would resolve nicely with the tonality of the verse.

\begin{figure}[htp!]\centering
    \includegraphics[width=.9\textwidth]{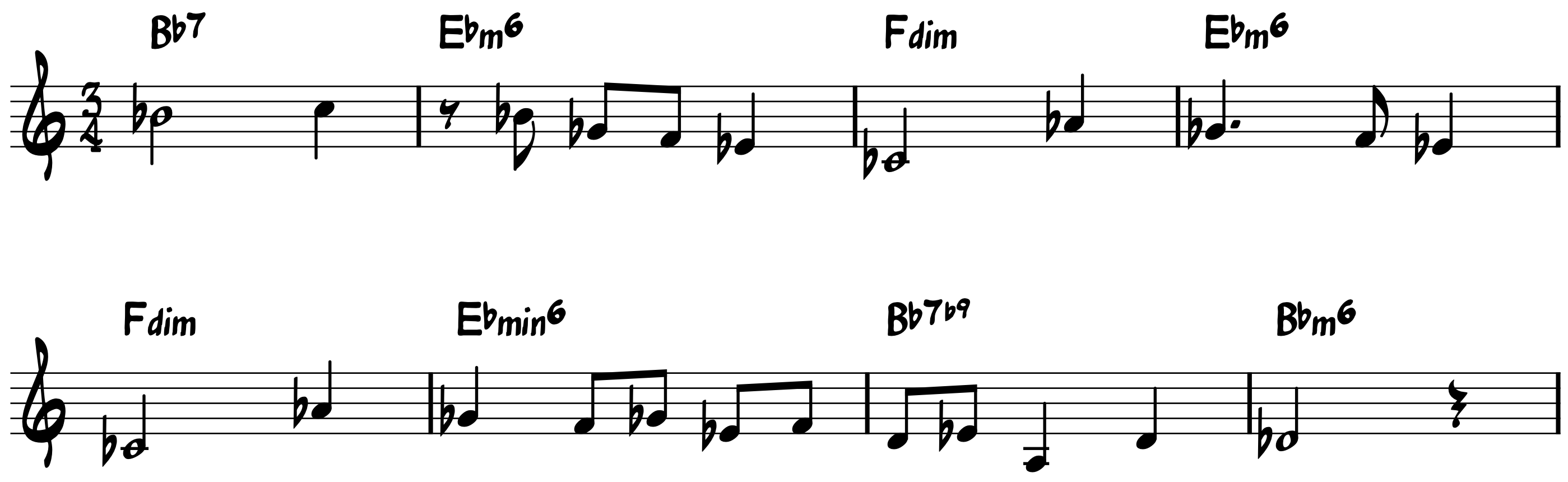}
    \caption{The catchy and singular verse of \emph{In the house of poetry}.}
    \label{fig:HouseOfPoetry1}
\end{figure}

He subsequently asked Kyrie Kristmanson to join, hoping she would like the song and would not be afraid of its technically challenging nature. She was indeed enthusiastic and wrote lyrics inspired by the tale \emph{The Shadow} by Andersen. She focused on the part of the story where the Shadow tells the learned man what he saw in the house of Poetry.

The song is divided in two parts. In the first part Kyrie sings; in the second part, Kyrie’s vocals are generated by Flow Machines from recordings of Kyrie’s voice.

\begin{figure}[htp!]\centering
    \includegraphics[width=.9\textwidth]{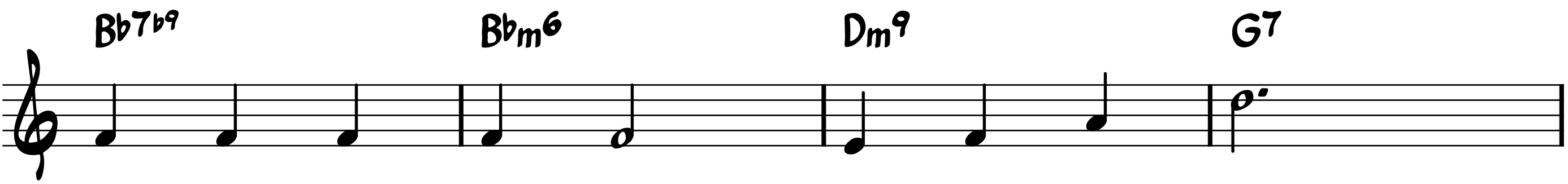}
    \caption{The chorus of \emph{In the house of poetry} features an unconventional and audacious ascending melody.}
    \label{fig:HouseOfPoetry2}
\end{figure}

Flow Machines generated pianos, strings and vocals from SKYGGE’s material as well as from Kyrie’s voice. SKYGGE played all additional instruments (drums, piano, and electric guitars).

\item \href{https://skyggewithai.bandcamp.com/track/cryyyy}{\emph{Cryyyy}}

This simple melody in the style of pop tunes form the 60s was generated almost exactly in its final form. When SKYGGE was working in the studio of Ash Workman with Michael Lovett, Ash had just received an old cabinet organ from the 70s bought on the net. They plugged it in and began to play the chords of \emph{Cryyyy}. It sounded great, and they recorded all those sounds for the song.
SKYGGE wanted a melancholic but modern sound and the timbre of Mariama’s voice matched perfectly. Flutes and detuned and distorted guitars were generated by Flow Machines, and SKYGGE added some beats and deep bass.

\item \href{https://skyggewithai.bandcamp.com/track/hello-shadow}{\emph{Hello Shadow}}

Stromae was fascinated by the possibilities of the software, and he fed the machines with his own influences: scores and audio stems in the Cape Verdian style. He selected his favorite melodies and stems from what Flow Machines had generated. Those fragments were put together and the song was built step by step. Stromae sung a vocal line that followed the generated melody, and he improvised on the pre-chorus. The choir in the chorus was also generated. When the song was ready for final production, it was sent  to singer Kiesza who loved it. She wrote lyrics inspired by the tale \emph{The Shadow}. Kiesza envisioned a happy, shiny shadow.  A most unusual and characteristic feature of this song are the first four notes of the verse, which evoke the image of a ball bouncing and rolling.

\item \href{https://skyggewithai.bandcamp.com/track/mafia-love}{\emph{Mafia Love (a.k.a. Oyestershell)}}

This song was composed by SKYGGE with Flow Machines, inspired by pop of the 60. It is almost a direct composition by Flow Machines, with very little human edits to the melody. The song has its internal logic, like all good songs. It tells a story, though unconventionally due to its rich structure, with almost no repetition. This absence of repetition sounds seem strange at first, but the song becomes an ear worm after hearing it a couple of times. The song was rendered with a generated voice from an a cappella recording of Curtis Clarke Jr., and a generated piano track from a stem by SKYGGE. 

\item \href{https://skyggewithai.bandcamp.com/track/paper-skin-feat-jata}{\emph{Paper Skin}}

Paper Skin is an interesting, indirect use of AI. This song was built from the song \emph{Mafia Love} (see above). JATA picked up fragments of \emph{Mafia Love} for the verse and the pre-chorus. He then composed a new chorus fitting those fragments. Ash Workman added some sounds from \emph{Mafia Love} in the intro and in the bridge. \emph{Paper Skin} is an offshoot of \emph{Mafia Love}, illustrating how a melodic line can travel ears and be transformed according to unpredictable inspirations of musicians.

\item \href{https://skyggewithai.bandcamp.com/track/multi-mega-fortune-feat-michael-lovett}{\emph{Multi Mega Fortune}}

Michael Lovett from NZCA Lines fed Flow Machines with his own audio stems, vocals, drums loops, bass and keyboards, as well as lead sheets in the style of Brit pop. A lot of material was generated, both songs and stems. Michael Lovett and SKYGGE curated the results, and Lovett wrote lyrics inspired by the tale The Shadow. The result is a synth-pop catchy tune with a distinctive gimmick generated by Flow Machines, that runs throughout the song.

\item \href{https://skyggewithai.bandcamp.com/track/valise-feat-the-pirouettes}{\emph{Valise}}

When Stromae came to the studio we tried several ideas based on six lead sheets generated with Flow Machines. Between sessions, SKYGGE explored one of those directions and generated a vocal line from the song \emph{Tous les m\^emes}, a former song of Stromae, uploaded on Flow Machines. The lyrics produced from the generated vocals meant something different from the original song by Stromae, but they were relevant, since they addressed the theme of ``luggage'' (Valise, in French). Stromae liked the song, but at the time we focused on the song \emph{Hello Shadow} and left \emph{Valise} aside for a while. SKYGGE asked the French band ``The Pirouettes'' to sing the melody instead of using the generated voice. The Pirouettes sung in sync with the generated voice by Stromae.
This song bears an uncommon yet catchy chord progression and structure. Like other songs from the album, first hearings may sound strange, but after a couple of listenings, the song becomes an ear worm. One can hear the generated choir laughing on top of this unsual chord progression.

\item \href{https://skyggewithai.bandcamp.com/track/cake-woman-feat-camille-bertault-et-mederic-collignon}{\emph{Cake Woman}}

M\'ed\'eric Collignon, an amazing jazz trumpet player, came to the lab  full of energy, with his own audio tracks, mostly jazz progressions played on a Rhodes piano, and some bass synths. SKYGGE also brought some hip-hop grooves, and the two musicians worked on a funk pop in the style of the 80s.
The generated lead sheet was simple but contained harmonic twists that M\'ed\'eric digs as a jazz composer. He selected chromatic modulations that he often uses in his own scores.

When they generated audio stems for the song from all the audio material, the output was messy but sounded very exciting. In particular, the groovy Rhodes generated from M\'ed\'eric’s own recordings reinforced the funkiness of the song. Pachet and SKYGGE wrote lyrics inspired by the nonsensical words of the generated voice, in the style of surrealist poetry. They asked the young and talented jazz singer Camille Bertault to sing the song. She also performed a scat-like improvisation, echoing the trumpet solo. In Andersen’s tale, the Shadow is hiding in the coat of a ``cake woman'', hence the title.

\item \href{https://skyggewithai.bandcamp.com/track/whistle-theme-feat-generated-vocals-from-catastrophe}{\emph{Whistle Theme}}

Themes from older soundtracks are often more melodic than recent ones. Today, film scores are more often based on textures than melodies. Inspired by those old soundtracks, Flow Machines generated a catchy theme for this song (see Figure~\ref{fig:WhistleTheme.png}). The whistling was backed up by airports sounds generated to match the song. Pierre Jouan from the pop band ``Catastrophe'' sung another song, from which the machine generated the voice heard in the song.

\begin{figure}[htp!]\centering
    \includegraphics[width=1\textwidth]{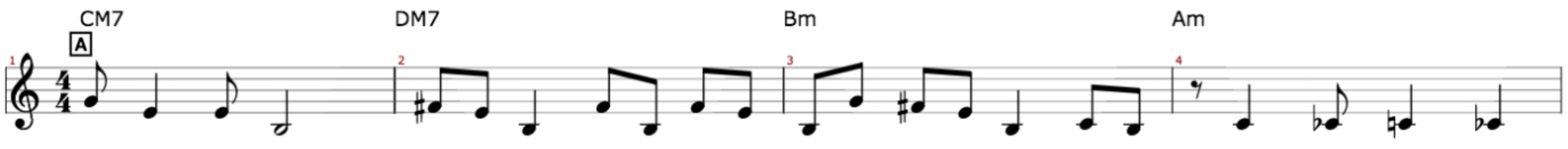}
    \caption{The catchy line of \emph{Whistle song}.}
    \label{fig:WhistleTheme.png}
\end{figure}

\item \href{https://skyggewithai.bandcamp.com/track/je-vais-te-manger-feat-sarah-yu-zeebroeke-laurent-bardainne}{\emph{Je Vais Te Manger}}

Laurent Bardainne came to the studio with his audio stems, marimba, synth bass patterns and compositions in the style of 80's pop. Flow Machines generated a few songs, and Laurent selected the good parts. Laurent and SKYGGE built the song in a few hours. They left each other without knowing what to do with their song. For over a week, SKYGGE woke up every morning with this melody stuck in his head. Looking in depth at the generated lead sheet, one can see a harmonic twist that no one would have thought of. This twist pushes the melody up and down over an audacious modulation (see~Figure\ref{fig:JeVaisTeManger}). SKYGGE and Laurent asked Sarah Yu to sing. The song is about a woman who says she will eat our souls and that we are lost.

\begin{figure}[htp!]\centering
    \includegraphics[width=1\textwidth]{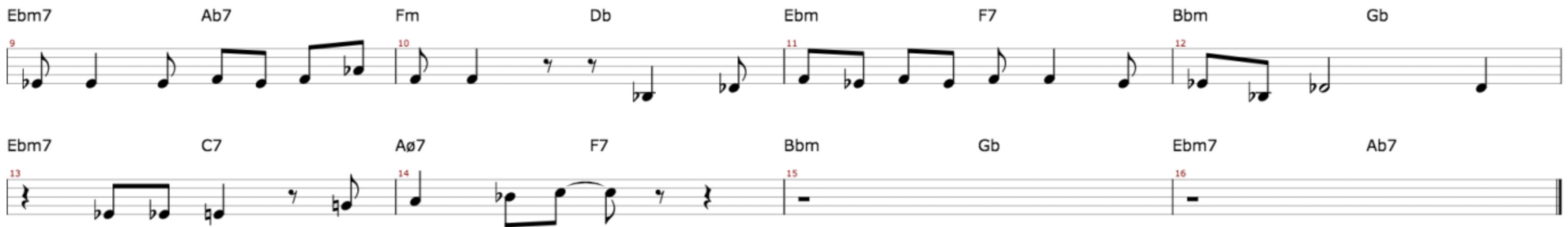}
    \caption{The catchy line of \emph{Je Vais Te Manger}.}
    \label{fig:JeVaisTeManger}
\end{figure}

\item \href{https://skyggewithai.bandcamp.com/track/cold-song}{\emph{Cold Song}}

The \emph{Cold Song} is a well-known part of the opera \emph{King Arthur} by Purcell. It has been sung by many singers in many styles (Sting, Klaus Nomi). For this cover, SKYGGE was inspired by artists such as Andre Bratten, Anne Clarke and Johann Johannsson, who have used  machines to produce melancholic moods. The voice is generated from an a cappella recording of singer Kyte. It turns out that the generation produced many “A I A I”, by coincidence. In this song cover, everything was produced by Flow Machines, and there was no manual production.

\end{enumerate}

\item \emph{Hello World Remix}

A few months after the release of \emph{Hello World}, a \href{https://skyggewithai.bandcamp.com/album/hello-world-remixed}{remix album} was released~\cite{HelloWorldRemixSpotify}, with remixes of 10 songs, produced by various producers. This album did not involve the use of AI, but is an important sign that the original songs of \emph{Hello World} were deemed sufficiently rich musically by several top remixers to deserve a remix. 
\end{enumerate}

\section{Unfinished but promising projects}
\label{section:Unfinished}

The Flow Machines project generated a rich flow of ideas and experiments. Some of these ideas did not contribute directly to music making, but are worth mentioning, to illustrate how an ambitious goal supported by a generous grant (the ERC) can lead to very high levels of creativity and productivity.

\begin{enumerate}

\item Continuator II

The Continuator system, which was at the root of the Flow Machines project, was extended, improved, and tested, notably during the Miror project. Extensions consisted mostly to add a constraining features, notably positional, as described in Section~\ref{section:positionalConstraints}, thereby improving substantially the musicality of the responses. Many experiments involving children and the question of music education based on reflexive interactions were conducted and described in~\cite{MirorBook}. A composition software (Miror-Compo) was implemented, allowing users too build fully fledged musical compositions that would reuse their doodlings (see Figure~\ref{fig:MirorCompo}).

\begin{figure}[htp!]\centering
    \includegraphics[width=.9\textwidth]{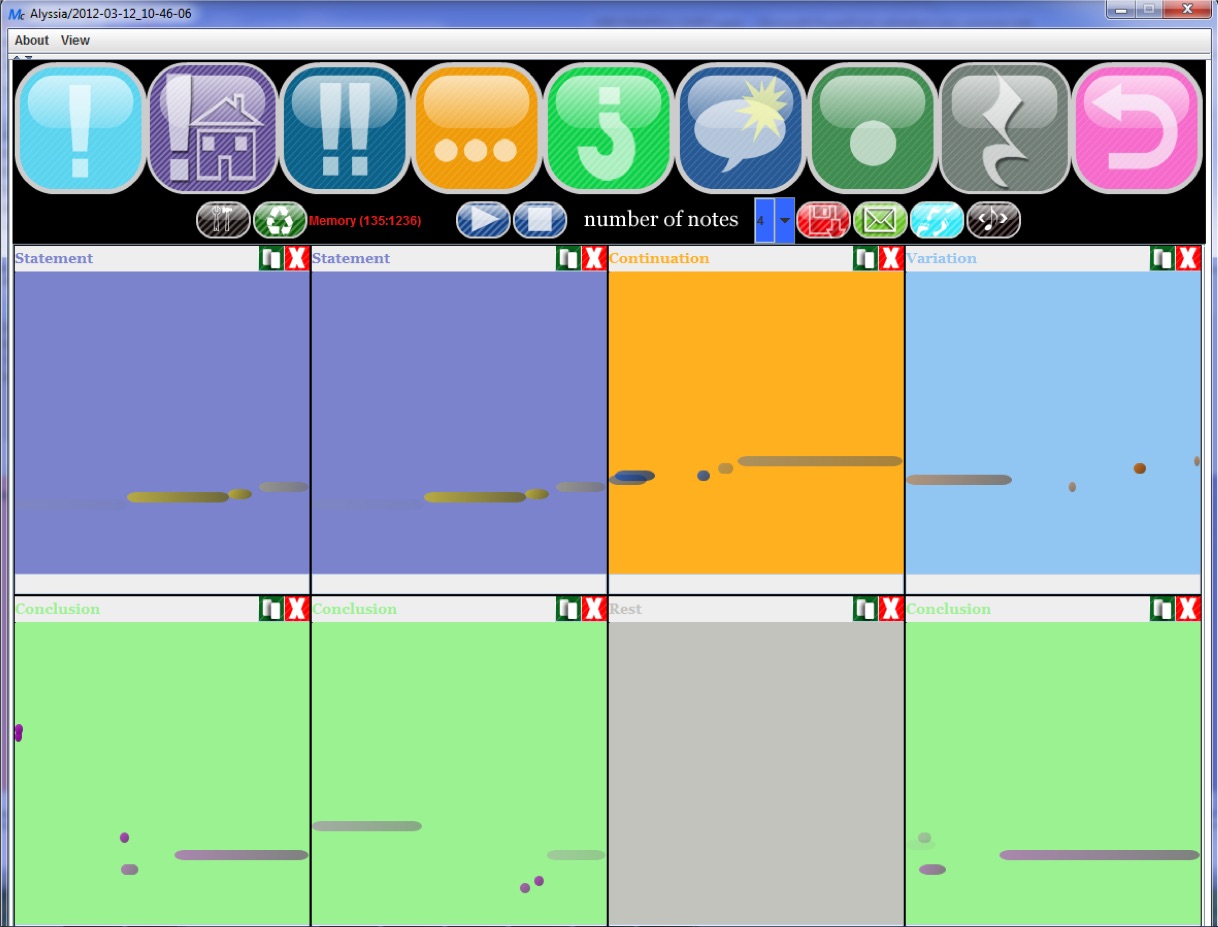}
    \caption{A composition software for children, that allows the to build pieces by reusing their previous doodlings.}
    \label{fig:MirorCompo}
\end{figure}

\item Film Music

Experiments in Film Music by Pierre Roy consisted in implementing the theory proposed in~\cite{murphy2014transformational,MurphyVideo}. The basic idea of Murphy is to consider all pairs of consecutive chords, and assign them an emotional category. It turns out that enforcing such typical pairs of chords can be approximated by biases of binary factors of the belief propagation algorithm. For instance, sadness is related to the pattern \emph{M4m}, \ie a major chord, followed by a minor chord 4 semitones (2 tones) above. Figure~\ref{fig:BeatlesSadness} shows lead sheet a generated in the style of Bill Evans (for the melody) and the Beatles (for the chords) biased to favor chord pairs related to ``sadness''~\cite{BeatlesSadnessCompo}.

\begin{figure}[htp!]\centering
    \includegraphics[width=.9\textwidth]{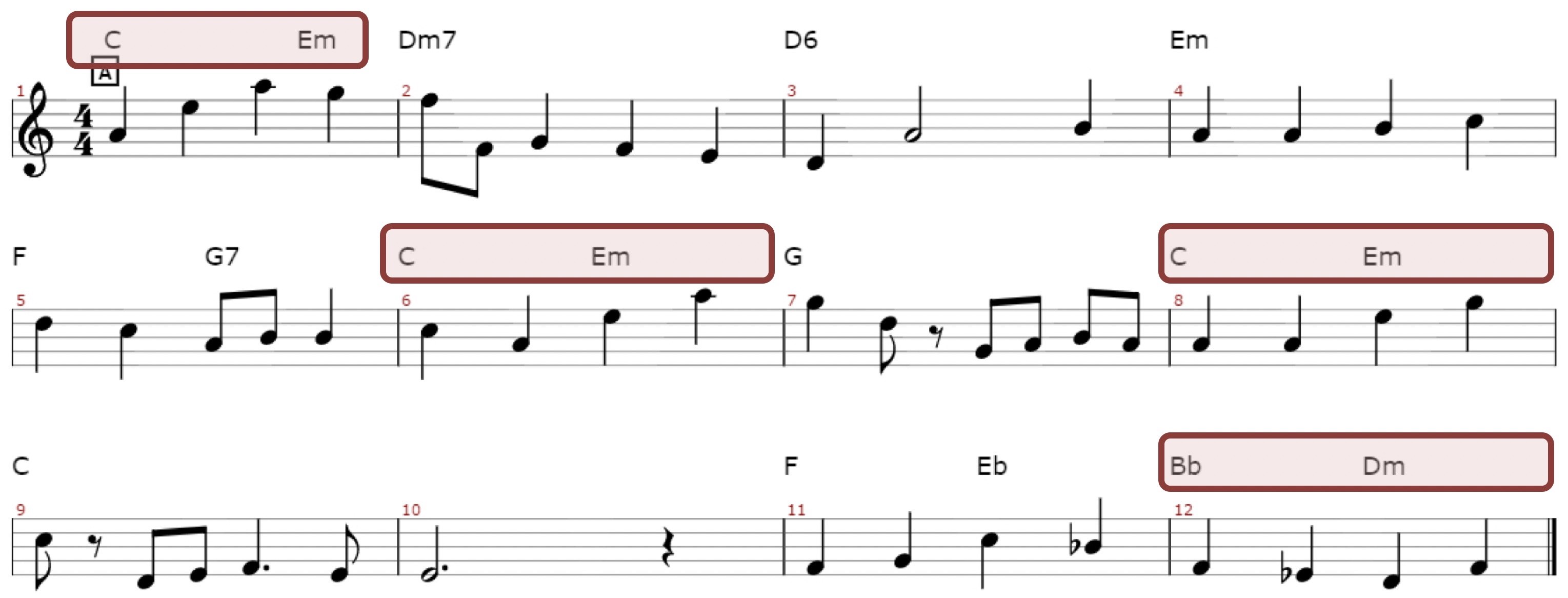}
    \caption{A lead sheet in the style of the Beatles, with a bias towards ``sadness'' related chord pairs: C to Em (3 times), Bb to Dm.}
    \label{fig:BeatlesSadness}
\end{figure}

\item Reflexive Looper

Reflexive Looper is the latest instantiation of a project consisting in building an interactive, real-system dedicated to solo performance. The initial fantasy was that of a guitarist playing in solo, but in need of a bass and harmony backing that would be tightly coupled to his real time performance. The core ideas of the system are described in~\cite{DBLP:conf/chi/PachetRMd13,DBLP:conf/cmmr/FoulonRP13} and many experiments and refactors were done during several years~\cite{DBLP:conf/ismir/MoreiraRP13}, culminating by a version implemented by Marco Marchini~\cite{Marchini2017RethinkingRL}. This system was nominated for the Guthman Instrument Competition and was demonstrated to a jury including guitarist Pat Metheny.

\item Interaction studies in jazz improvisation

A question that arose regularly in our work about real time assisted improvization was how to model actual humans interacting togeter when they improvize. It turns out that little is known about the nature of such interactions. In particular, a key question is to which extent this interaction involves the content of the music (rhythm, harmony, melody, expressiveness)? Such a question was crucial for designing of our music interaction systems. We proposed in~\cite{pachet2017jazz} an analytical framework to identify correlates of content-based interaction. We illustrated the approach with the analysis of interaction in a typical jazz quintet (the Mark d'Inverno quintet). We extracted audio features from the signals of the soloist and the rhythm section. We measured the dependency between those time series with correlation, cross-correlation, mutual information, and Granger causality, both when musicians played concomitantly and when they did not. We identified a significant amount of dependency, but we showed that this dependency was mostly due to the use of a common musical context, which we called the \emph{score effect}. We finally argued that either content-based interaction in jazz is a myth or that interactions do take place but at unknown musical dimensions. In short, we did not see any real interaction!

\item Feedback in lead sheet composition

We studied the impact of giving and receiving feedback during the process of lead sheet composition in~\cite{MartinFeedback}. To what extent can peer feedback affect the quality of a music composition? How does musical experience influence the quality of feedback during the song composition process? Participants composed short songs using an online lead sheet editor, and are given the possibility to provide feedback on other participants’ songs. Feedback could either be accepted or rejected in a later step. Quantitative data were collected from this experiment to estimate the relation between the intrinsic quality of songs (estimated by peer evaluation) and the nature of feedback. Results show that peer feedback could indeed improve both the quality of a song composition and the composer’s satisfaction about it. Also, it showed that composers tend to prefer compositions from other musicians with similar musical experience levels.

\item Experiments in collective creativity (a.k.a the rabbits experiment)

In order to better understand the impact of feedback and collective creation, we designed an experiment to highlight different types of creator profiles~\cite{DBLP:conf/ijcai/GhediniFPR15}. In this experiment, users could write captions to short comic strips, were shown what other users in a group would do, and given the possibility to ``switch'' to the productions of their peers. One of the conclusions is that the potential impact of implicit feedback from other participants and objectivity in self-evaluation, even if encouraged, are lessened by a bias “against change”. Such a bias probably stems from a combination of self-enhancing bias and of a commitment effect.

\item A Comic book on style creation

A comic book, titled \emph{MaxOrder}, was designed and produced, telling a story about a character (Max) who struggles to invent her own style~\cite{DBLP:conf/ijcai/GhediniPR15}. The name of the character stemmed from the MaxOrder result, which concerns the limitation of plagiarism in Markov sequence generation~\cite{papadopoulos2016generating,DBLP:conf/aaai/PapadopoulosRP14}. The comic, available on the web~\cite{MaxOrderWeb}, gave rise to the ERCcOMICS~\cite{ErcComicsSite} project, which produced comics about 16 ERC projects.

\begin{figure}[htp!]\centering
    \includegraphics[width=.9\textwidth]{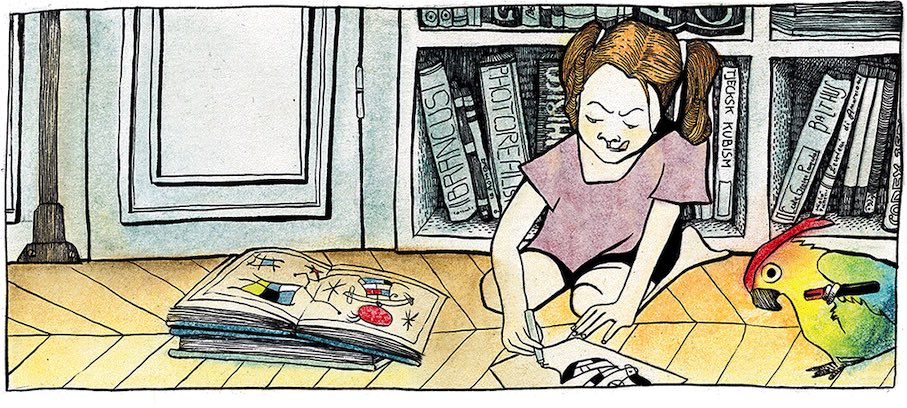}
    \caption{The character named MaxOrder, who tries to create her own style.}
    \label{fig:MaxOrder}
\end{figure}

\item Song composition project

In this project, the author participated actively in the composition and release of 2 albums: one in Pop music (\emph{Marie-Claire}~\cite{MarieClaireAlbum}) and one in jazz (\emph{Count on it}~\cite{CountOnItAlbum}). Most composition sessions were recorded on video for later analysis. The jazz album was premiered in a sold-out concert at Pizza club express, one of the main jazz venue in London.

\item Automatic Accompaniment generation

Following the Flow Composer project (see above), we addressed the issue of real time accompaniments using similar techniques, \ie based on belief propagation. We recorded jazz pianist Ray d'Inverno (see Figure~\ref{fig:RaydInverno}) as well as saxophonist Gilad Atzmon, and produced various accompaniment algorithms (listen for instance to a \href{https://www.francoispachet.fr/wp-content/uploads/2020/04/Girl_from_Ipanema_mix.mp3}{compelling accompaniment} of \emph{Girl From Ipanema} produced using an accompaniment on \emph{Giant Steps}~\cite{GeneratedAccompaniment2017}).

\begin{figure}[htp!]\centering
    \includegraphics[width=.9\textwidth]{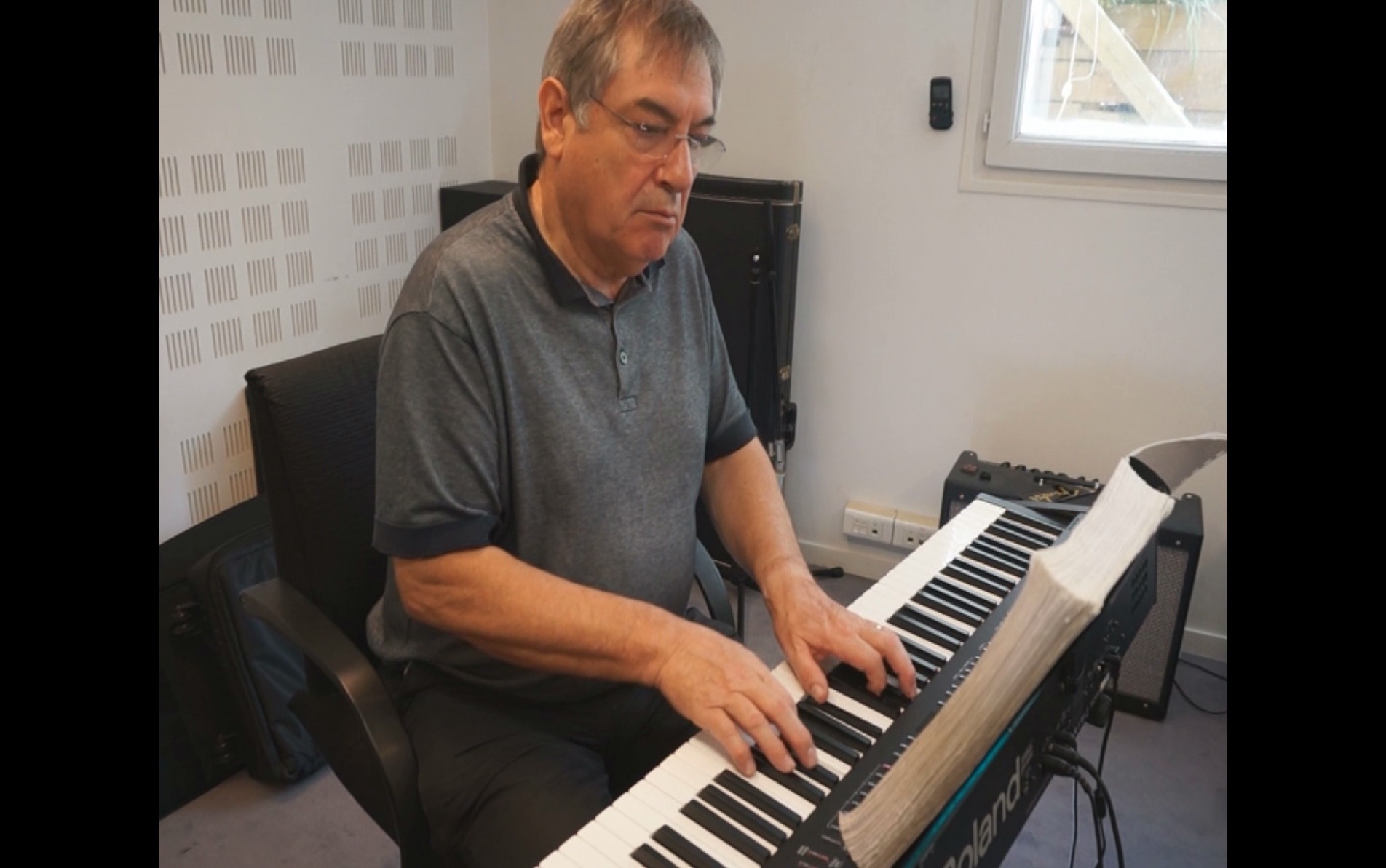}
    \caption{Jazz pianist Ray d'Inverno recording an accompaniment on \emph{Girl from Ipanema}.}
    \label{fig:RaydInverno}
\end{figure}

\item Interview series

Several \href{https://www.francoispachet.fr/music-style-interviews/}{interviews of composers} were conducted during the project: Ennio Morricone, \href{https://www.youtube.com/watch?list=PLvoqwxjRRNfmVSPoRDUkXfr6KkB7gimV1&v=GL0rjJ-y9e0&feature=emb_title}{Ivan Lins}, \href{https://www.youtube.com/watch?list=PLvoqwxjRRNfnC12o8fBsnbv1ulivocAyg&v=GJSdFze_76I&feature=emb_title}{Benoit Carr\'e}, \href{https://www.youtube.com/watch?list=PLvoqwxjRRNfmSVdDAQQRSK0OenL9IGkpn&time_continue=1&v=k3WTF59_ew4&feature=emb_title}{Hubert Mounier (L'affaire Louis Trio)}, Michel Cywie (composer of beautiful French pop songs in the 70s), \href{https://www.youtube.com/watch?list=PLvoqwxjRRNfk2fDi9aCMgM828PhiUozJB&v=MORzAzeR7KY&feature=emb_title}{Franco Fabbri}. The initial idea was to track down what composers of famous hits thought of the song writing process, to form a book about how songwriters make hits.

\item Flow-Machines radio

A prototype of an online web radio was developed. This web generated on-the-fly pieces composed by Flow Composer, rendered them with our concatenative synthesis system, and streamed them on a dedicated web site. Users could rate the songs they listened. The goal was to obtain enough feedback on songs to use reinforcement learning to tune automatically the generators (see Figure~\ref{fig:FMRadio}). 

\begin{figure}[htp!]\centering
    \includegraphics[width=.8\textwidth]{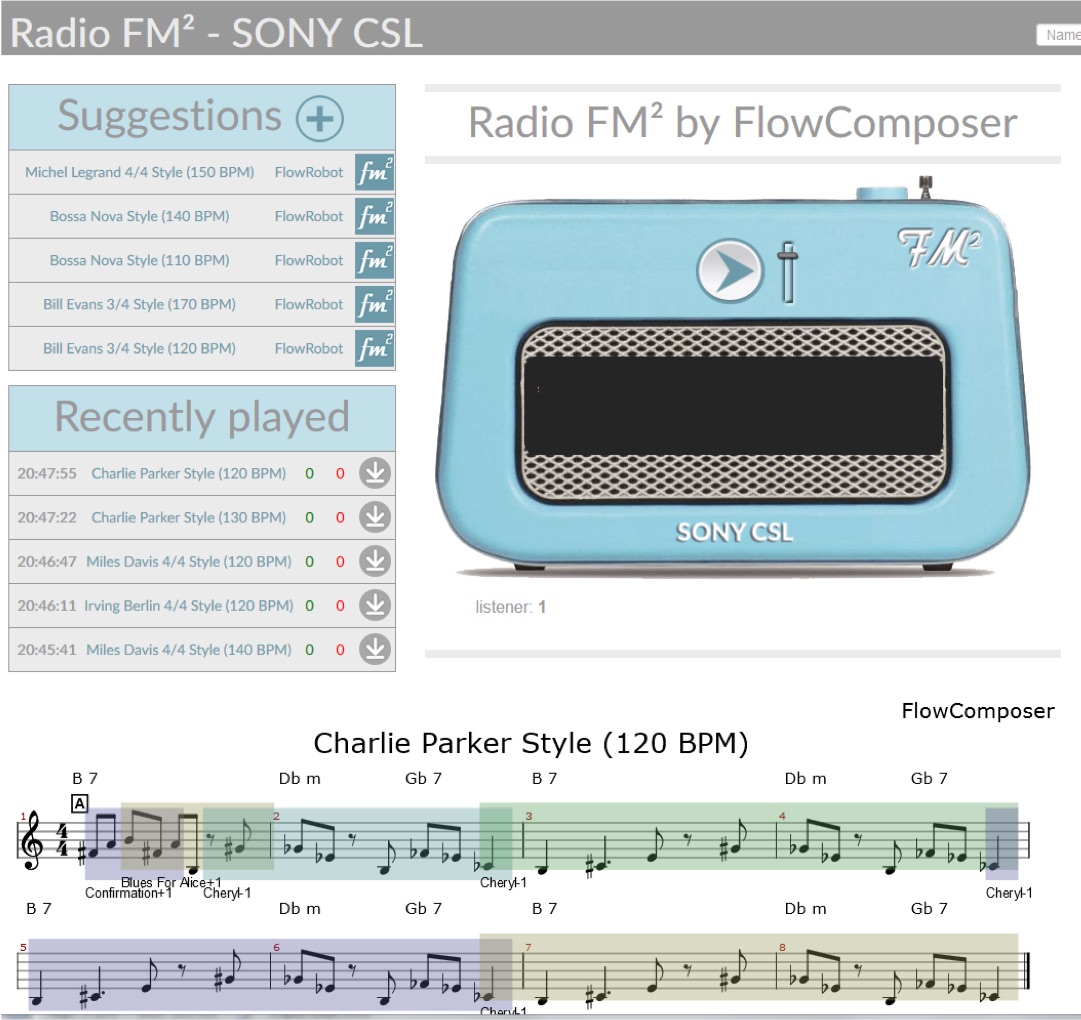}
    \caption{A snapshot of a web radio broadcasting automatically generated music, with user feedback.}
    \label{fig:FMRadio}
\end{figure}

\end{enumerate}

\section{Impact and followup}

Three years after the completion of the project, what remains the most important for us in these developments is not the technologies we built \emph{per se}, though many strong results were obtained in the \emph{Markov + X} roadmap, but the music that was produced with these technologies. In that respect, the utmost validation of this work lies in the following:

\begin{enumerate}
\item The overwhelming media reception of the album composed with it~\cite{HelloWorldMedia}, including some outstanding reviews, such as: ``AI and humans collaborated to produce this hauntingly catchy pop music''~\cite{AvdeefQZ}, or ``Is this the world's first good robot album?''~\cite{MarshallBBC}
\item The enthusiasm of all the musicians who participated in this projet. 15 songs were composed and signed by various artists (including Stromae), and this is in itself a validation, as dedicated musicians would not sign songs they are not proud of 
\item The overall success of the album on streaming platforms (a total of 12 million streams as of April 2020), with \emph{Magic Man} (5.9 million streams) and \emph{Hello Shadow} (2.8 million) as main hits  
\item The positive reception by music critics. For instance, Avdeef writes~\cite{Avdeef}:
\begin{quote}SKYGGE’s \emph{Hello World} is a product of these new forms of production and consumption, and functions as a pivot moment in the understanding and value of human-computer collaborations. The album is aptly named, as it alludes to the first words any new programmer uses when learning to code, as well as serving as an introduction to new AI-human collaborative practices. Hello, World, welcome to the new era of popular music.
\end{quote}
Similarly, emphasising the difference between interactive AI-assisted song composition, which Flow Composer pioneered, and fully automatic composition, Miller writes~\cite{Miller}: 
\begin{quote}
On the one hand, we have Fran\c{c}ois Pachet’s Flow Machines, loaded with software to produce sumptuous original melodies, including a well-reviewed album. On the other, researchers at Google use artificial neural networks to produce music unaided. But at the moment their music tends to lose momentum after only a minute or so.
\end{quote}
\item Sony Flow Machines

The Flow Machines project is continuing at Sony CSL~\cite{FlowMachinesNew,FlowMachinesCSL}, and research around the ideas developed in the project has been extended, for instance with new interactive editors in the same vein as Flow Composer~\cite{DBLP:journals/corr/abs-1907-10380}. Promising musical projects have also been setup, notably involving Jean-Michel Jarre~\cite{FlowMachinesJarre}. 

\item Beyond Flow Machines

Since \emph{Hello World}, SKYGGE has launched another album made with Artificial Intelligence, \href{https://skyggewithai.bandcamp.com/album/american-folk-songs}{\emph{American Folk Songs}}~\cite{AmericanFolkSongsAlbum}. In this album, original \emph{a cappellas} of traditional songs (notably by Pete Seeger, with his right owner's approval) were reorchestrated with automatic orchestration tools developed at Spotify.  The techniques used are based on a belief propagation scheme~\cite{OrchestratorPaper} combined with advanced edition features~\cite{DBLP:journals/corr/abs-1903-08459}. All the orchestrations were generated by \textcolor{red}{orchestration transfer}, \ie transferring the orchestration style of existing music pieces to these traditional folk melodies. Ongoing work to use large-scale listening data (skip profiles) to tune the model are ongoing with promising results~\cite{DBLP:journals/corr/abs-1903-06008}.

\end{enumerate}

\section{Lessons learned}

\subsection{Better model does not imply better music}
We experimented with various generative models: variable-order Markov models, max entropy and deep learning techniques. 
Paradoxically, the most interesting music generation were not obtained with the most sophisticated techniques.
In spite of many attempts to use deep learning methods upfront for generation, we did not succeed in using these techniques for professional use (at the time of the project). The Bach chorales are of undisputed high quality, but they are not very \emph{interesting} musically: they produce either ``correct'' material, or mistakes, but not much in between.
However, we observed that the combination of various tools (\eg, the lead sheet generation tool and our audio synthesis tool) produced flow states more easily than the use of single algorithms. This may have to do with the appropriation effect mentioned below.

\subsection{New creative acts}

The use of generative algorithms introduces new tasks for the user, and these tasks are actually creative acts, in the sense that they require some musical expertise or, at least intuition and intention: 1) the selection of training sets and 2) the curation of the results. Of course, training set selection and curation can be performed using random algorithms, but so far we did not succeed in automating theses processes.

\subsection{The appropriation effect}

More importantly, there seems to be a tradeoff to find between the sophistication of the algorithm and the sense of appropriation of the results by the user. 
The IKEA effect is a cognitive bias which has been confirmed by many experiments, in which consumers place a disproportionately high value on products they partially created~\cite{Landers2012}. Many consumer studies have shown for instance that instant cake mixes were more popular when the consumer had something to do (such as adding an egg) than when the instant mix was self sufficient. It is likely that in the case of AI assisted creation, the same type of effect applies, and that users do not get a sense of appropriation if the algorithm, whatever its sophistication, does all the creative job. This is a fascinating area of study that will surely produce counter-intuitive results in the future.

\section{Toward new Categories of \emph{new}}

The traditional conceptual landscape used to describe novelty in music is based essentially on three notions: 1) \emph{original} songs (new material, possibly inspired by preexisting work, but to an acceptable degree), 2) \emph{covers} (new orchestrations of existing songs), 3) \emph{plagiaristic work}, \ie works containing segments of existing songs with no or little transformations. Another category is \emph{parasiting}, the action of imitating an orchestration style without plagiarising it, in order to avoid paying royalties~\cite{LeTavernier16}, but this category is apparently less used by music professionals. 
A lot of arguments used in the debate about the role of AI in music creation touch on the notion of creativity or novelty, \ie can AI produce really novel music material. However, we dont think the question ``is AI creative'' is relevant, since we see AI as a tool for creators, an opinion shared by most researchers in the field (\eg~\cite{Fiebrink2016TheML}).
Yet, we argue the idea that all the generations we described here, and the ones to come, are instances of  new categories of ``new''.
In this text, we have described several pieces generated with various techniques, and used intentionally the following terms (in red):

\begin{enumerate}
\item \textcolor{red}{stylistic explorations}: Song snippets ``in the style of'', the song \emph{Sensitive}
\item \textcolor{red}{stylistic singularities}: the \emph{Boulez Blues}
\item \textcolor{red}{reminiscences}: \emph{Scratch my itch}
\item \textcolor{red}{pastiches and exercises}: \emph{Daddy's car} and Bach chorales
\item \textcolor{red}{orchestration transfers}: the orchestrations in the album \emph{American Folk Songs}, which were all produced by transferring existing orchestration styles to existing songs
\item \textcolor{red}{timbre transfers}: most of the renderings of the examples described here
\item \textcolor{red}{templagiarism}: the song \emph{Lunar}
\end{enumerate}

It is our view that these new categories may help us conceptualize the contributions of these generative technologies to music creation. Interestingly, images created with Deep Dream~\cite{45507} and referred to as \textcolor{red}{hallucinations} do not have their equivalent yet in music: what would be, indeed a musical hallucination ?

\section{Conclusion}

We have described a number of research results in the domain of assisted music creation, together with what we consider are remarkable music productions. We attempted to describe why these generations are interesting, as technological artefacts, and also from a musical viewpoint. Since the Flow Machines project ended (2017), research in computer music generation has exploded, mostly due to the progress in deep learning~\cite{briot2019deep}. After the launch of \emph{Daddy's car} and \emph{Hello World}, several music titles were produced with AI (notably by artists Taryn Southern and Holly Herndon), contributing to the exploration of these technologies for music creation.

These explorations are still only scratching the surface of what is possible with AI. However, they already challenge the status and value of what is considered ``new'' in our digital cultures. We have sketched a draft of a vocabulary to describe and distinguish different type of ``newness'' in music. These categories start to have well defined meaning technically but much more is needed to give them precise definitions. Surely, more categories will arise, some will vanish. In the end, only music and words will remain, not technologies.

\section{Acknowledgements}
The Flow Machines project  received funding from the European Research Council under the European Union’s Seventh Framework Programme (FP/2007-2013) / ERC Grant Agreement n. 291156.
We thank the team of the musical \emph{Beyond the Fence} for their insightful comments in using earlier versions of Flow Composer. We thank the researchers and engineers who participated in the Flow Machines project: Gabriele Barbieri, Vincent Degroote, Gaëtan Hadjeres, Marco Marchini, Dani Mart\'in, Julian Moreira, Alexandre Papadopoulos, Mathieu Ramona, Jason Sakellariou, Emmanuel Deruty and Fiammetta Ghedini. We also thank professors Jean-Pierre Briot (CNRS), Mirko Degli Esposti (University of Bologna), Mark d'Inverno (Goldsmiths College, University of London), Jean-Fran\c{c}ois Perrot (Sorbonne Universit\'e) and Luc Steels  (VUB), for their precious and friendly contributions. We thank our colleagues and friends Giordano Cabral, Geber Ramalho, Jean-Charles R\'egin, Guillaume Perez, Shai Newman for their insights. We also thank all the musicians who were involved in the projects, under the supervision of Benoit Carr\'e:  Stromae, Busy P, Catastrophe, ALB, Kumisolo, Olivier Marguerit, Barbara Carlotti, Housse de Racket, Camille Bertault, Lescop, Raphael Chassin, Michael Lovett, Ash Workman, C Duncan, Marie-Jeanne Serrero, Fred Dec\'es, Mariama, Kyrie Kristmanson, The Bionix, Kiesza, JATA, The Pirouettes, M\'ed\'eric Collignon, Pierre Jouan, Sarah Yu Zeebroeke, Laurent Bardainne. Finally, we thank our host institutions Sony CSL, UPMC, Spotify, the label Flow Records and the distributor Idol for hosting, supporting and believing in this research.

%
%
\bibliographystyle{splncs04}
\bibliography{biblio,dl4mg}

\end{document}